\documentclass[fleqn,usenatbib]{mnras}
\DeclareRobustCommand{\ion}[2]{%
\relax\ifmmode
\ifx\testbx\f@series
{\mathbf{#1\,\mathsc{#2}}}\else
{\mathrm{#1\,\mathsc{#2}}}\fi
\else\textup{#1\,{\mdseries\textsc{#2}}}%
\fi}

\usepackage{newtxtext}
\usepackage{gensymb}
\usepackage{supertabular}
\usepackage[T1]{fontenc}
\usepackage[slantedGreek]{newtxmath}

\DeclareRobustCommand{\VAN}[3]{#2}
\let\VANthebibliography\thebibliography
\def\thebibliography{\DeclareRobustCommand{\VAN}[3]{##3}\VANthebibliography}

\usepackage{graphicx}	% Including figure files
\usepackage{amsmath}	% Advanced maths commands

\title[AT 2023sva]{Multi-Wavelength Analysis of AT 2023sva: a Luminous Orphan Afterglow With Evidence for a Structured Jet}

\author[G. P. Srinivasaragavan et al.]{Gokul P. Srinivasaragavan$^{1,2,3}$,
Daniel A. Perley$^{4}$,
Anna Y. Q. Ho$^5$,
Brendan O'Connor$^{6,7}$,
\newauthor 
Antonio de Ugarte Postigo$^{8,9}$,
Nikhil Sarin$^{10,11}$,
S. Bradley Cenko$^{2,3}$,
Jesper Sollerman$^{12}$,
\newauthor
Lauren Rhodes$^{13,14}$,
David A. Green$^{15}$,
Dmitry S. Svinkin$^{16}$,
Varun Bhalerao$^{17}$,
Gaurav Waratkar$^{17}$,
\newauthor
A.J. Nayana$^{18}$,
Poonam Chandra$^{19, 18}$,
M. Coleman Miller$^{1,2}$,
Daniele B. Malesani$^{20, 21, 22}$,
Geoffrey Ryan$^{23}$,
\newauthor
Suryansh Srijan$^{24}$,
Eric C. Bellm$^{25}$,
Eric Burns$^{26}$,
David J. Titterington$^{15}$,
Maria B. Stone$^{27}$,
\newauthor
Josiah Purdum$^{28}$,
Tomás Ahumada$^{29}$,
G.C. Anupama$^{30}$,
Sudhanshu Barway$^{30}$,
Michael W. Coughlin$^{31}$,
\newauthor
Andrew Drake$^{29}$,
Rob Fender$^{32}$,
Jos\'e F. Ag\"u\'i Fern\'andez$^{33}$,
Dmitry D. Frederiks$^{16}$,
\newauthor
Stefan Geier$^{34,35}$,
Matthew J. Graham$^{29}$, 
Mansi M. Kasliwal$^{29}$,
S. R. Kulkarni$^{29}$,
Harsh Kumar$^{36}$,
\newauthor
Maggie L. Li$^{29}$,
Russ R. Laher$^{37}$,
Alexandra L. Lysenko$^{16}$,
Gopal Parwani$^{17}$,
Richard A. Perley$^{38}$,
\newauthor
Anna V. Ridnaia$^{16}$,
Anirudh Salgundi$^{17}$,
Roger Smith$^{37}$,
Niharika Sravan$^{39}$,
Vishwajeet Swain$^{17}$,
\newauthor
Christina C. Th\"{o}ne$^{40}$,
Anastasia E. Tsvetkova$^{41,16}$,
Mikhail V. Ulanov$^{16}$,
Jada Vail$^{5}$,
Jacob L. Wise$^{4}$,
\newauthor
Avery Wold$^{37}$
\\
$^{1}$Department of Astronomy, University of Maryland, College Park, MD 20742, USA\\
$^{2}$Joint Space-Science Institute, University of Maryland, College Park, MD 20742, USA\\
$^{3}$Astrophysics Science Division, NASA Goddard Space Flight Center, 8800 Greenbelt Rd, Greenbelt, MD 20771, USA\\
$^{4}$Astrophysics Research Institute, Liverpool John Moores University, 146 Brownlow Hill, Liverpool L3 5RF, UK\\
$^{5}$Department of Astronomy, Cornell University, Ithaca, NY 14853, USA\\
$^{6}$McWilliams Fellow\\
$^{7}$McWilliams Center for Cosmology and Astrophysics, Department of Physics, Carnegie Mellon University, Pittsburgh, PA 15213, USA\\
$^{8}$Universit\'e de la C\^ote d’Azur, Observatoire de la C\^ote d’Azur, CNRS, Artemis, Nice, F-06304, France\\
$^{9}$Aix Marseille Univ, CNRS, LAM, Marseille, France\\
$^{10}$Oskar Klein Centre for Cosmoparticle Physics, Department of Physics, Stockholm University, AlbaNova, Stockholm SE-106 91, Sweden\\
$^{11}$Nordita, Stockholm University and KTH Royal Institute of Technology, Hannes Alfvens vag 12, SE-106 91 Stockholm, Sweden\\
$^{12}$Oskar Klein Centre, Department of Astronomy, Stockholm University, AlbaNova, SE-10691 Stockholm, Sweden\\
$^{13}$Trottier Space Institute at McGill, 3550 Rue University, Montreal, Quebec H3A 2A7, Canada\\
$^{14}$Department of Physics, McGill University, 3600 Rue University, Montreal, Quebec H3A 2T8, Canada\\
$^{15}$ Astrophysics Group, Cavendish Laboratory, 19 J J Thomson Avenue,
Cambridge CB3 0HE, UK\\
$^{16}$Ioffe Institute, Polytekhnicheskaya, 26, St. Petersburg, 194021, Russia\\
$^{17}$Department of Physics, IIT Bombay, Powai, Mumbai 400076, India\\
$^{18}$National Centre for Radio Astrophysics, Tata Institute of Fundamental Research, Pune University Campus, Ganeshkhind Pune 411007, India\\
$^{19}$National Radio Astronomy Observatory, 520 Edgemont Rd, Charlottesville VA 22903\\
$^{20}$Cosmic Dawn Center (DAWN), Denmark\\
$^{21}$Niels Bohr Institute, University of Copenhagen, Jagtvej 128, Copenhagen, 2200, Denmark\\
$^{22}$Department of Astrophysics/IMAPP, Radboud University Nijmegen, P.O. Box 9010, Nijmegen, 6500 GL, The Netherlands\\
$^{23}$Perimeter Institute for Theoretical Physics, Waterloo, Ontario N2L 2Y5, Canada\\
$^{24}$Department of Computer Science and Engineering, IIT Bombay, Powai, Mumbai 400076, India\\
$^{25}$DIRAC Institute, Department of Astronomy, University of Washington, 3910 15th Avenue NE, Seattle, WA 98195, USA\\
$^{26}$Department of Physics and Astronomy, Louisiana State University, Baton Rouge, LA 70803 USA\\
$^{27}$Department of Physics and Astronomy, Vesilinnantie 5, FI-20014, University of Turku, Finland\\
$^{28}$Caltech Optical Observatories, California Institute of Technology, Pasadena, CA 91125, USA\\
$^{29}$Division of Physics, Mathematics, and Astronomy, California Institute of Technology, Pasadena, CA 91125, USA\\
$^{30}$Indian Institute of Astrophysics, 2nd Block 100 Feet Rd, Koramangala Bangalore, 560034, India\\
$^{31}$School of Physics and Astronomy, University of Minnesota,
Minneapolis, MN 55455, USA\\
$^{32}$Astrophysics, Department of Physics, University of Oxford, Keble Road, Oxford OX1 3RH, UK\\
$^{33}$Centro Astron\'omico Hispano en Andaluc\'ia, Observatorio de Calar Alto, Sierra de los Filabres, G\'ergal, Almer\'ia, 04550, Spain\\
$^{34}$Cuesta de San Jos\'{e} s/n, 38712 Bre{\~n}a Baja, La Palma, Spain\\
$^{35}$Instituto de Astrof\'{\i}sica de Canarias, V\'{\i}a   L\'actea, 38205 La Laguna, Tenerife, Spain\\
$^{36}$Center for Astrophysics | Harvard \& Smithsonian, 60 Garden St. Cambridge MA, 02138, USA\\
$^{37}$IPAC, California Institute of Technology, 1200 E. California Blvd, Pasadena, CA 91125, USA\\
$^{38}$National Radio Astronomy Observatory, P.O. Box ‘O’, Socorro, NM 87801\\
$^{39}$Department of Physics, Drexel University, Philadelphia, PA 19104, USA\\
$^{40}$Astronomical Institute, Czech Academy of Sciences, Fric\v ova 298, Ond\v rejov, Czech Republic\\
$^{41}$Dipartimento di Fisica, Università degli Studi di Cagliari, SP
Monserrato-Sestu, km 0.7, I-09042 Monserrato, Italy\\
}

\date{Accepted XXX. Received YYY; in original form ZZZ}

\pubyear{2024}

\begin{document}
\label{firstpage}
%\pagerange{\pageref{firstpage}--\pageref{lastpage}}

\maketitle

\clearpage

% These dates will be filled out by the publisher

% Abstract of the paper
\begin{abstract}
We present multi-wavelength analysis of ZTF23abelseb (AT 2023sva), an optically discovered fast-fading ($\Delta m_r =  2.2$ mag in $\Delta t = 0.74 $ days), luminous ($M_r \sim -30.0$ mag) and red ($g-r = 0.50$ mag) transient at $z = 2.28$ with accompanying luminous radio emission. AT 2023sva does not possess a $\gamma$-ray burst (GRB) counterpart to an isotropic equivalent energy limit of $E_{\rm{\gamma, \, iso}} < 1.6 \times 10^{52}$ erg, determined through searching $\gamma$-ray satellite archives between the last non-detection and first detection, making it the sixth example of an optically-discovered afterglow with a redshift measurement and no detected GRB counterpart. We analyze AT 2023sva's optical, radio, and X-ray observations to characterize the source. From radio analyses, we find the clear presence of strong interstellar scintillation (ISS) 72 days after the initial explosion, allowing us to place constraints on the source's angular size and bulk Lorentz factor. When comparing the source sizes derived from ISS of orphan events to those of the classical GRB population, we find orphan events have statistically smaller source sizes. We also utilize Bayesian techniques to model the multi-wavelength afterglow. Within this framework, we find evidence that AT 2023sva possesses a shallow power-law structured jet viewed slightly off-axis ($\theta_{\rm{v}} = 0.07 \pm 0.02$) just outside of the jet's core opening angle ($\theta_{\rm{c}} = 0.06 \pm 0.02$). We determine this is likely the reason for the lack of a detected GRB counterpart, but also investigate other scenarios. AT 2023sva's evidence for possessing a structured jet stresses the importance of broadening orphan afterglow search strategies to a diverse range of GRB jet angular energy profiles, to maximize the return of future optical surveys.
\end{abstract}

\section{Introduction}
\label{Intro}

A small subset of stripped-envelope core-collapse supernova (CCSN) explosions are accompanied by Long Gamma-ray Bursts (LGRBs; duration $T_{90} > 2 \, \rm{s}$; \citealt{Cano2017}). LGRBs are powered by accretion onto a black hole remnant or the rotational spin-down of a neutron star remnant, generating collimated ($\theta_0 \approx 10 \degree $) ultra-relativistic ($\Gamma > 100$) jets \citep{Macfadyen1999}. Traditionally, emission from GRBs is divided into two phases -- the prompt emission in $\gamma$-rays originates from within the jets and the afterglow emission across the electromagnetic spectrum originates from the interaction of these jets with the surrounding medium \citep{vanparadijs2000, Panaitescu2000}.

There have been hundreds of optical afterglows detected through follow-up observations of well-localized LGRB triggers,
%\footnote{An up-to-date list is maintained at \url{http://www.mpe.mpg.de/∼jcg/grbgen.html}}, 
along with over 50 of their associated supernovae (see, e.g., \citealt{Galama1998,Hjorth2003,Hjorth2013, Cano2017,Melandri2019, Hu2021, Kumar2022, Rossi2022, Blanchard2023, Srinivasaragavan2023, Srinivasaragavan2024, Finneran2024}). The advent of state-of-the art time domain surveys including the Zwicky Transient Facility (ZTF; \citealt{Graham2019, Bellm2019, Dekany2020, Masci2019}) have also enabled the serendipitous discovery of optical afterglows without an associated GRB trigger. Searches of GRB archives post-facto to several optical afterglow discoveries have shown a number of events possess associated GRBs not discovered by high-energy satellites \citep{Cenko2015, Bhalerao2017, Stalder2017, Melandri2019, Ho2022}. Possible explanations for these optical discoveries include GRBs producing $\gamma$-ray emission that do not notify GRB satellites to send out prompt alerts, or the lack of a robust afterglow localization. 
%uncovered a new class of optically discovered GRB-related phenomena known as ``orphan afterglows". These transients possess  -- all features shared with GRB optical afterglows. However, they are discovered independently of GRB triggers, and their origins are mysterious.

If post-facto searches through GRB archives do not find observed associated $\gamma$-ray emission, these afterglows are known as  
``orphan" afterglows\footnote{The term ``apparently" orphan is a more precise term to describe these events, as this is a purely observational definition. Physically, an associated GRB to some isotropic energy limit usually cannot be ruled out for optically-discovered afterglows, and it is possible that $\gamma$-ray emission may be present in these systems, but was just not observed. For brevity, we refer to these events purely as ``orphan" for the rest of the text. }. These orphan afterglows may arise from a few different scenarios. The
simplest explanation is that GRB satellites may have missed the prompt $\gamma$-ray emission due to limited coverage in certain regions of the sky, or being turned off due to operational reasons. Another explanation is due to extremely off-axis classical GRBs. Since GRBs are ultra-relativistic, emission from the jet at early times is only observable within a viewing angle of $\theta \sim 1/\Gamma$. As the jet slows down, the relativistic beaming cone widens \citep{Rhoads1997, Meszaros1998} and by the time the afterglow generates emission at optical wavelengths, the cone could include Earth's line of sight. 
%Therefore, it is possible that optically-discovered afterglows may be off-axis classical LGRBs where GRB satellites are not in the line of sight of the initial jets. 
 The invocation of a ``structured" jet is also a  possible explanation, where a GRB's energy profile varies with respect to viewing angle \citep{Gottlieb2021, Gottlieb2022, Granot2010}. For a structured jet, even slightly off-axis observers viewing an event within the jet's viewing angle but outside the jet's narrow high-$\Gamma$ core would see an orphan afterglow \citep{ Nakar2003,Rossi2008, Cenko2013,Salafia2015, Gavin2017,Gavin2018, Huang2020, Sarin2021, OConnor+2023, Freeburn2024}, sometimes dubbed an ``on-axis" orphan \citep{Nakar2003}.

 Another possibility is that a GRB's jet is extremely baryon-loaded, reducing its Lorentz factor. The higher density of baryons can result in pair production processes absorbing $\gamma$-ray prompt emission, reradiating it at longer wavelengths. These baryon-loaded LGRBs have been proposed as ``dirty fireballs" \citep{Dermer1999}, though there have been no observationally confirmed dirty fireballs discovered thus far in the literature. The discovery of even one genuine dirty fireball would change our picture of GRB phenomena, confirming long-held theories that baryon-loaded jets can successfully break out of their progenitor stars \citep{Paczynski1998,Dermer1999}. Some other proposed scenarios include stalled, choked jets forming a cocoon of shocked material that produces little to no $\gamma$-rays \citep{Gottlieb2018}, with the interaction of the cocoon with the surrounding medium producing an afterglow similar to classical GRB jets at a lower luminosity \citep{Nakar2017}, and low radiative efficiency bursts \citep{Sarin2022}.

The study of optically-discovered afterglows started only around 14 years ago. The first ever discovery was in 2011, PTF11agg \citep{Cenko2013}. The event did not have a confirmed redshift, though it was argued to be between $z = 1$ and $z = 2$. Follow-up observations using the Jansky Very Large Array (VLA; \citealt{Perley2011}) showed that there was a long-lived scintillating radio counterpart to the optical transient. Two other optically-discovered afterglows (iPTF14yb and ATLAS17aeu) were discovered in a similar manner to PTF11agg at extragalactic distances \citep{Cenko2015, Bhalerao2017, Stalder2017, Melandri2019}. However, searches through GRB archives after their discoveries showed that both events had associated observed $\gamma$-ray emission that high-energy satellites did not promptly send out notifications for regarding their discovery.

The discovery space for these events changed dramatically with ZTF, as its rapid near-nightly cadence and wide field of view made it a prime instrument for discovering optical afterglows serendipitously. Through dedicated afterglow searches, 11 ZTF-discovered afterglows have been published since ZTF's inception in March 2018 \citep{Ho2020c, Ho2022, Andreoni2021, Andreoni2022, Perley2024, Li2024}. Nine of these events have confirmed redshift measurements through optical spectroscopy and five have no associated GRB found post-facto, making them orphan events (AT 2019pim, AT 2020blt, AT 2021any, AT 2021lfa, and AT 2023lcr). 

%A redshift measurement is crucial to determine whether the non-detection of an associated GRB can rule out the prescence of a normal, on-axis classical GRB. AT2020blt, AT2021any, and AT2021lfa all possessed high enough redshifts such that limits on accompanying $\gamma$-ray emission were not able to rule out the presence of an on-axis classical GRB. However, AT2019pim was the first example of an orphan afterglow where an on-axis classical GRB was able to be ruled out with certainty \citep{Perley2024}. 

Because there have been so few orphan afterglow discoveries, studies probing the physical origins of newly discovered events are important for understanding their nature. Though optical observations are utilized for their discovery, understanding their full physical picture necessitates follow-up observations in the X-ray and radio wavelengths, where jet physics and ejecta characteristics can be probed. Indeed, modeling of AT 2021any's X-ray through radio emission suggests a possible low-Lorentz factor origin \citep{Xu2023} or a classical GRB missed by high-energy satellites \citep{Gupta2022, Li2024}, AT 2020blt and AT 2023lcr were best modeled as classical GRBs missed by high-energy satellites \citep{Ho2020c, Li2024}, and AT 2019pim and AT 2021lfa were best modeled as originating from either low-Lorentz factor GRBs or slightly off-axis structured jet GRBs \citep{Perley2024, Li2024, Lipunov2022}.

In this work, we present the optical, radio, and X-ray observations of an orphan afterglow, ZTF23abelseb (AT 2023sva) at $z = 2.28$, making it the sixth such event presented in the literature.  We utilize AT 2023sva's multi-wavelength observations to physically characterize the source. The paper is organized as follows: in \S \ref{Observations} we present optical, X-ray, and radio observations of AT 2023sva, in \S \ref{opradioanalysis} we analyze the multi-wavelength data set, in \S \ref{interpret} we provide a physical interpretation of the afterglow, and in \S \ref{conclusion} we summarize our results and present conclusions. We note that throughout this paper we utilize a flat $\Lambda$CDM cosmology with $\Omega_{\rm m}=0.315$ and $H_{0} = 67.4$~km~s$^{-1}$~Mpc$^{-1}$ \citep{Planck18} to convert the redshift to a luminosity distance and correct for the Milky Way extinction of $E(B-V)_{\rm{MW}} = 0.24$ \citep{Schlafly2011}, and host galaxy extinction of $E(B-V)_{\rm{host}} = 0.09$ mag (see \S \ref{SEDoptical}). 

\section{Observations}
\label{Observations}
\subsection{ZTF Discovery}
\label{ZTFdisc}

AT2023sva was discovered by ZTF \citep{GCNdisc} at $r = 17.71 \pm 0.05$ mag (all magnitudes are in the AB system), on 2023-09-17 09:38:31.20 (all times are given in UTC), at a location $\alpha$ (J2000)= 00$^{\mathrm{h}}$56$^{\mathrm{m}}$59\fs20$^{\mathrm{s}}$,
$\delta$ (J2000) = +80$^{\circ}$08$\arcmin44\farcs13$. ZTF is a survey on the 48-inch telescope at Palomar Observatory that covers around 10000 deg$^2$ every night \citep{Bellm2019}, enabling it to survey the entire northern sky in the $g$ and $r$ bands every two nights, along with $i$ band for certain pre-selected fields. The survey's observing system is described in \citet{Dekany2020} and transient discovery utilizes an image subtraction pipeline \citep{Zackay2016} utilizing deep reference images of fields \citep{Masci2019}. We use the ZTF Fritz marshal to store the photometry \citep{VanderWalt2019, Coughlin2023skyportal}.

The line of sight MW extinction is $E(B-V)_{\rm{MW}} = 0.24$, corresponding to $A_V = 0.74$ mag. Correcting for MW extinction and the host-galaxy extinction derived in \S \ref{SEDoptical}, at a redshift of $z = 2.28$ (see \S \ref{spectroscopy}), this first detection corresponds to an absolute magnitude in $r$ band of $M_r \sim -30.0$ mag, making AT 2023sva an extremely luminous optical transient. The source was flagged by human scanners after passing a filter in the ZTF alert stream \citep{patterson2019} that searches for young and fast transients (described in \citealt{Ho2020c}). The source was not detected to a limiting magnitude of $r > 20.36$ mag two nights before on 2023-09-15 05:50:11.95, implying a rapid rise rate of $>$ 1.3 mag day$^{-1}$. There was no host galaxy counterpart detected in the ZTF reference images (the deepest upper limit was $g > 20.7$ mag) nor in PanSTARRS images of the field. The source also decayed rapidly, at a rate of 3 mag day$^{-1}$ in the $r$ band after the initial observations.

Optically-discovered afterglows are classified and differentiated from false positives by their rapid rises and decays, red colors indicative of a synchrotron spectrum, extragalactic redshifts, and extremely high luminosities (a full description of how optical afterglows are discovered in ZTF's alert stream is presented in \citealt{Ho2020c}). The primary false positives in optically-discovered afterglow searches are stellar flares in the Milky Way. These flares possess blackbody temperatures of around 10,000 K \citep{Kowalski2013}, and in the optical bands, their spectrum lies on the Rayleigh--Jeans tail. This corresponds to a spectrum with $f_\nu \propto \nu^2$, or an extinction-corrected blue color of $g-r = -0.17$ mag. Afterglows, on the other hand, have characteristic red colors due to their synchrotron spectrum, with $f_\nu \propto \nu^{-\beta}$ where $f_\nu$ is the flux density and $\beta$ is the spectral index, the exponential factor that relates the flux density of a source to its frequency. A $g$-band observation was obtained of AT 2023sva shortly after the initial $r$-band detection, of $g = 18.52 \pm 0.02$, on 2023-09-17 10:41:57.034. Extrapolating this $g$-band detection to the time of the first r-band detection assuming a simple power-law evolution (detailed in \S \ref{opticalLCsection}), the extinction corrected $g-r$ color at the time of discovery is $\sim 0.5$ mag. The fast-rise, red colors, and lack of a host galaxy counterpart made AT 2023sva an optically-discovered afterglow candidate and motivated follow-up observations.

\subsection{Follow-up Observations}
\subsubsection{Optical Photometry}
We triggered a Target of Opportunity (ToO) program on the Spectral Energy Distribution Machine (SEDM; \citealt{Blagorodnova2018,Rigualt2019}) on the automated 60-inch telescope at Palomar Observatory, to obtain imaging of AT 2023sva in the $g$, $r$, and $i$ bands. The images were processed utilizing a \texttt{Python}-based version of the \texttt{Fpipe} \citep{FrSo2016} pipeline. The pipeline includes photometric calibrations and image subtraction utilizing reference images from The Sloan Digital Sky Survey (SDSS; \citealt{SDSS}). SEDM obtained four epochs of photometry, from 2023-09-18 03:56:16.080 to 2023-09-19 03:57:27.101. These follow-up observations showed AT 2023sva's light curve (LC) rapidly decayed by 2.2 mag in $r$ band, in just 0.74 days (see Figure \ref{opticalLC}). 

We also triggered a ToO program on the 0.7-m GROWTH-India Telescope \citep[GIT;][]{2022AJ....164...90K} located at the Indian Astronomical Observatory (IAO), Hanle-Ladakh, to obtain an additional epoch of $g$ and $r$ band imaging of AT 2023sva. The observations began on 2023-09-19 17:06:25.574. We utilized the \texttt{ZOGY} algorithm-based \texttt{Python} pipeline to perform image subtraction on the images using PS1 templates and to obtain the final photometry. 

% NOT observations by DBM

A single ToO observation in the $g$ and $r$ bands was also secured with the 2.56-m Nordic Optical Telescope (NOT), located in the Canary Islands (Spain) equipped with the ALFOSC imager, on 2023-09-18 20:15:12.96. Data reduction was carried out following standard techniques and photometric calibration was computed against the Pan-STARRS catalog. The complete photometry obtained of AT 2023sva is presented in Table \ref{photometrytable}.

%We also retrieved the transient magnitudes from images obtained in the spectroscopy acquisition procedure (see Sect.~\ref{spectroscopy}) at the 10.6-m Gran Telescopio Canarias (GTC), equipped with the OSIRIS+ instrument.

\begin{table}
\caption{Optical photometry of AT 2023sva. The photometry in this table is not corrected for MW or host-galaxy extinction. }
\begin{tabular}{c|c|c|c|c}
\hline
\hline
MJD  & Filter & AB mag & Uncertainty & Facility \\
\hline
60204.40175 & \textit{r} & 17.71 & 0.05 & ZTF \\ 
        60204.40175 & \textit{r}  & 17.71 & 0.05 & ZTF \\ 
        60204.40175 & \textit{r}  & 17.68 & 0.02 & ZTF \\ 
        60204.42189 & \textit{r}  & 17.71 & 0.05 & ZTF\\ 
        60204.42189 & \textit{r}  & 17.71 & 0.05 & ZTF\\ 
        60204.42189 & \textit{r}  & 17.70 & 0.01 & ZTF\\ 
        60204.44532 & \textit{g}  & 18.45 & 0.06  & ZTF\\ 
        60204.44532 & \textit{g} & 18.39 & 0.02  & ZTF\\ 
        60204.4458 & \textit{g} & 18.52 & 0.02 & ZTF\\ 
        60204.4458 & \textit{g} & 18.49 & 0.06 & ZTF\\ 
        60205.16078 & \textit{g} & 20.33 & 0.07 & SEDM \\ 
        60205.16243 & \textit{r} & 19.89 & 0.07& SEDM \\ 
        60205.16408 & \textit{i} & 19.38 & 0.05& SEDM \\ 
        60205.30556 & \textit{g} & 20.41 & 0.08& SEDM \\ 
        60205.30721 & \textit{r} & 19.79 & 0.06& SEDM \\ 
        60205.30887 & \textit{i} & 19.59 & 0.09& SEDM \\ 
        60205.48597 & \textit{g} & 20.73 & 0.10 & SEDM\\ 
        60205.48761 & \textit{r} & 20.22 & 0.07& SEDM \\ 
        60205.48925 & \textit{i} & 19.72 & 0.07 & SEDM\\ 
        60205.8439 & \textit{r} & 20.61 & 0.07 & NOT\\ 
        60205.8477 & \textit{g} & 21.18 & 0.09 & NOT\\ 
        60206.16083 & \textit{g} & 21.20 & 0.12& SEDM\\
        60206.1649 & \textit{r} & 21.10 & 0.15& SEDM \\ 
        60206.7128 & \textit{r} & 21.89 & 0.07 & GIT\\ 
        60206.8192 & \textit{g} & 22.68 & 0.11 & GIT\\
        60208.3729 & \textit{r} & $> 22.12$ & -- & SEDM\\
        60208.3666 & \textit{g} & $> 22.13$ & -- & SEDM\\
\end{tabular}
\label{photometrytable}
\end{table}

\subsubsection{Optical Spectroscopy}
\label{spectroscopy}
We secured spectroscopy of AT2023sva using OSIRIS+ \citep{2000SPIE.4008..623C} on the 10.4-m Gran Telescopio Canarias (GTC) on 2023-09-19 03:55:51.011. The observation was carried out using the R1000B grism, which covers the wavelength range 3600--7800~\AA, and consisted of 3 exposures of 900~s each \citep{GCN.34740}. The slit position angle was set to parallactic to minimize differential slit losses.

The data were reduced using a self-developed pipeline based on IRAF routines. Data reduction included bias and response correction and wavelength calibrations using HgAr and Ne lamps, which were also used to do a 2D distortion correction. Cosmic rays were removed using the \texttt{lacos\_spec} routine \citep[][]{2001PASP..113.1420V}.
%The flux calibration was performed using as reference the spectrophotometric standard star G191B2B \citep[][]{1990AJ.....99.1621O}.
The 1D spectrum was obtained through optimal extraction \citep{1986PASP...98..609H}.

\subsubsection{X-ray Observations}
\label{XrayObs}
Observations of AT 2023sva with the  \textit{Neil Gehrels Swift Observatory} \citep{Gehrels_2004} X-ray Telescope \citep[XRT;][]{Burrows2005} began at 2023-09-21 18:00:00.000, through a ToO trigger submitted by our team after the optical afterglow discovery. Observations lasted for 7.5 ks after the initial trigger. The data were obtained in Photon Counting (PC) mode. The transient was not detected in the observations, up to a 0.3 -- 10 keV flux limit of $< 1.2 \times 10^{-13} \, \rm{erg \, cm^{-2} \, s^{-1}}$. We convert this integrated flux limit to a maximum estimated flux density limit at 5 keV (to use in \S \ref{Modeling} when we model the multi-wavelength data), through assuming a photon index of $\Gamma = 1.75$. We use this value, which corresponds to the optical spectral index of $\beta = 0.75$, because we determine that the X-ray and optical data likely do not lie on the same spectral segment in \S \ref{SEDoptical}. The true X-ray spectral index must be steeper than $\beta = 0.75$, so we use $\beta = 0.75$ conservatively to extrapolate the 0.3 -- 10 keV flux limit to the flux density limit at 5 keV, in order for the upper limit derived to be the maximum allowed. We derive a flux density upper limit at 5 keV of $< 8.65 \times 10^{-15} \, \rm{erg \, cm^{-2} \, s^{-1} \, \rm{keV}^{-1}}$. 

\begin{figure}
    \centering
\includegraphics[width=0.99\linewidth]{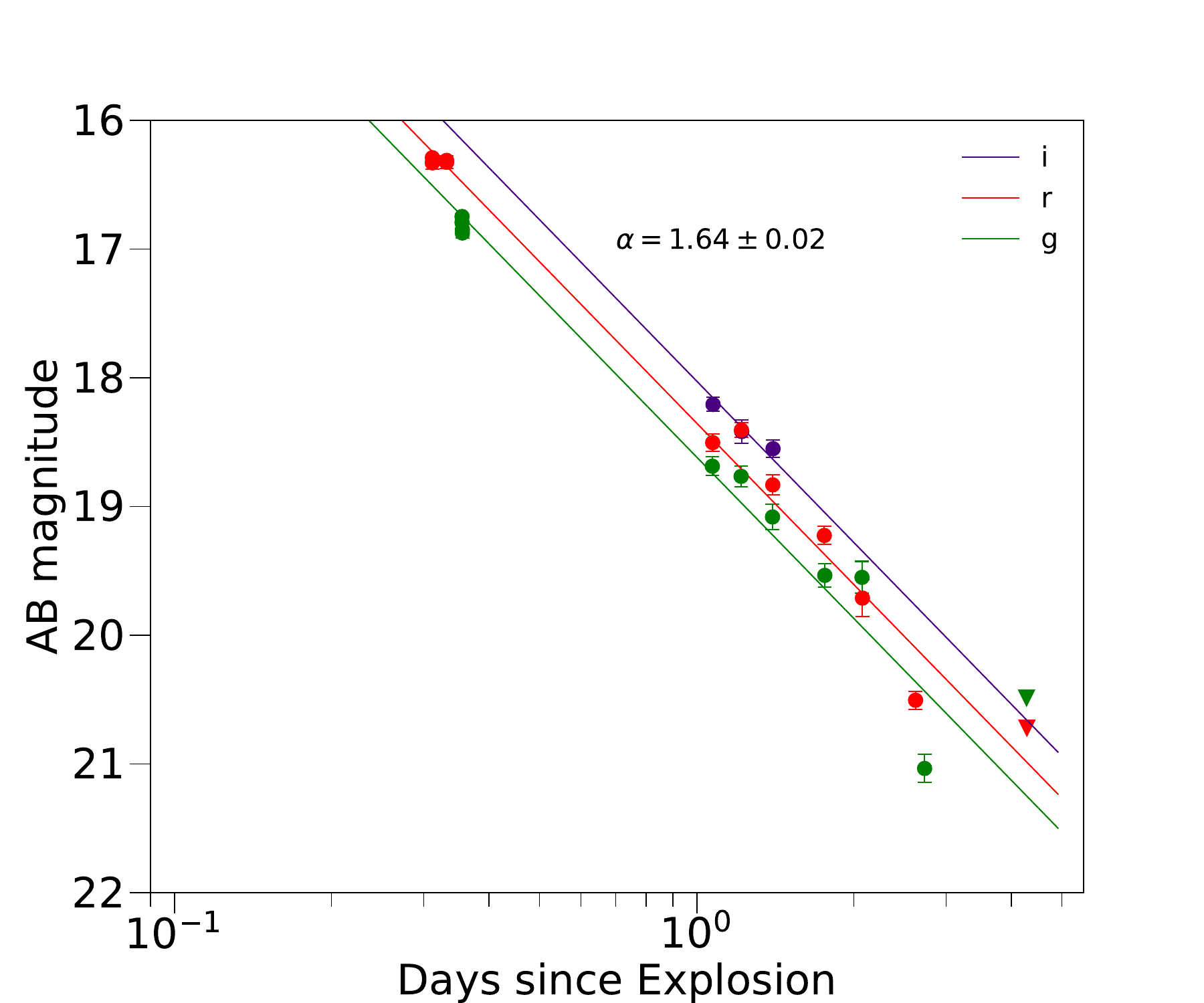}
    \caption{Optical LC of AT 2023sva in the $g$, $r$, and $i$ bands and the earliest $3\sigma$ upper limits, along with a best fit power-law decay model. The photometry is taken from Table \ref{photometrytable}, and then corrected for both MW and host-galaxy extinction.}
    \label{opticalLC}
\end{figure} 

\subsubsection{Radio Observations}

The Arcminute Microkelvin Imager – Large Array (hereafter AMI—LA) is a 8-dish
interferometer based at the Mullard Radio Astronomy Observatory outside
Cambridge in the UK. It observes at a central frequency of 15.5\,GHz with a
bandwidth of 5\,GHz \citep{2008MNRAS.391.1545Z, hickish2018}.
Observations of AT 2023sva commenced with AMI—LA on 2023-09-19 02:08:35UT (1.2\,days after the discovery). Each
observation consisted of a series of 600\,second scans of the target interleaved with
100\,second of the phase calibrator J0017+8135, adding up to a total of 4\,hours on
target (except on 2023-10-07 which was only
2\,hours). Once per day AMI—LA also observes a bandpass/flux calibrator 3C286.
AT 2023sva was observed over 12 epochs spanning 1 to 60\,days post-discovery.
Data from AMI—LA is reduced using custom software \textsc{reduce\_dc}
\citep{2013MNRAS.429.3330P} which performs flux scaling, bandpass and complex
gain calibration as well as flagging for radio frequency interference and antenna
shadowing. The calibrated data are then output in \textit{fits} format so they can be
read into \textsc{casa} for imaging and any further flagging that is required
\citep{ McMullin2007}. Imaging was performed interactively using \textit{tclean} within
\textsc{casa}. The AMI observations are presented in Table \ref{AMI}.
%The resulting flux densities and 3$\sigma$ upper limits are reported in Table \ref{tab:AMI}.

We also observed AT 2023sva with the Karl G. Jansky Very Large Array (VLA) on six occasions between November 2023 and August 2024 (program ID 24A-130).   The first observation used the S, C, X, Ku, and Ka-band receivers (3, 6, 10, 15, and 33 GHz); subsequent observations used only a subset of these receivers.  Data were calibrated and imaged using standard procedures in the Astronomical Image Processing System (AIPS). Flux density measurements were performed using \texttt{imfit}. The VLA observations are presented in Table \ref{VLA}.

\begin{figure}
    \centering
    \includegraphics[width=0.99\linewidth]{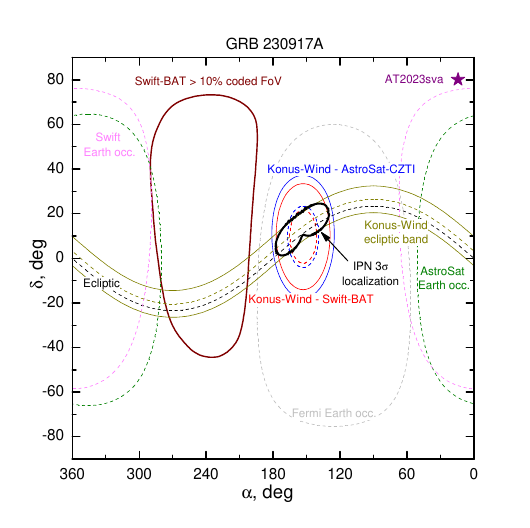}
    \caption{The IPN 3$\sigma$ localization region for GRB 230917A (the thick solid black line), compared to the point-source location of AT 2023sva (star). In addition, we show the Swift-BAT field of view (FOV; the solid white line), the Earth occulted regions for Swift, Fermi, and Astrosat-CZTI (dashed lines), the ecliptic plane (the dashed black line), the Konus-Wind ecliptic latitude constraint (the solid dark green lines indicates the 3$\sigma$ uncertainty and the dashed dark green line indicates the most probable burst source ecliptic latitude), the Konus-Wind -- AstroSat-CZTI and Konus-Wind -- Swift-BAT 3$\sigma$ triangulation annuli (solid blue and red lines), and the most probable burst location utilizing the Konus-Wind -- AstroSat-CZTI and Konus-Wind – Swift-BAT triangulation (dashed blue and red lines). We conclude that GRB 230917A is not associated with AT 2023sva.
}
    \label{IPNlocalization}
\end{figure}

%\textbf{Add uGMRT}
We also carried out Upgraded Giant Metrewave Radio Telescope (uGMRT) observations of AT\,2023sva at two epochs during 22-31 March and 17-20 May 2024. The data were recorded in standard continuum observing mode with a time integration of 10 seconds in bands-3 (250$-$500 MHz), 4 (550$-$850 MHz), and 5 (1000$-$1460 MHz). We used a bandwidth of 200 MHz in bands 3 and 4 and 400 MHz in band-5 split into 2048 channels. 3C48 was the flux density and bandpass calibrator and 2344+824 (bands-4 and 5) and 0229+777 (band-3) were the phase calibrators. We used Astronomical Image Processing Software \citep[AIPS;][]{Greisen2003} to reduce the data following standard procedure \citep{Nayana2017}. The data were initially inspected for non-working antennae and radio frequency interference. The corrupted data were flagged using tasks \texttt{UVFLG}, \texttt{TVFLG}, and \texttt{SPFLG} and then calibrated. The calibrated target data were imaged using task \texttt{IMAGR} in an interactive mode followed by a few rounds of phase-only self-calibration.  
\begin{figure*}
    \centering
    \includegraphics[width=0.99\linewidth]{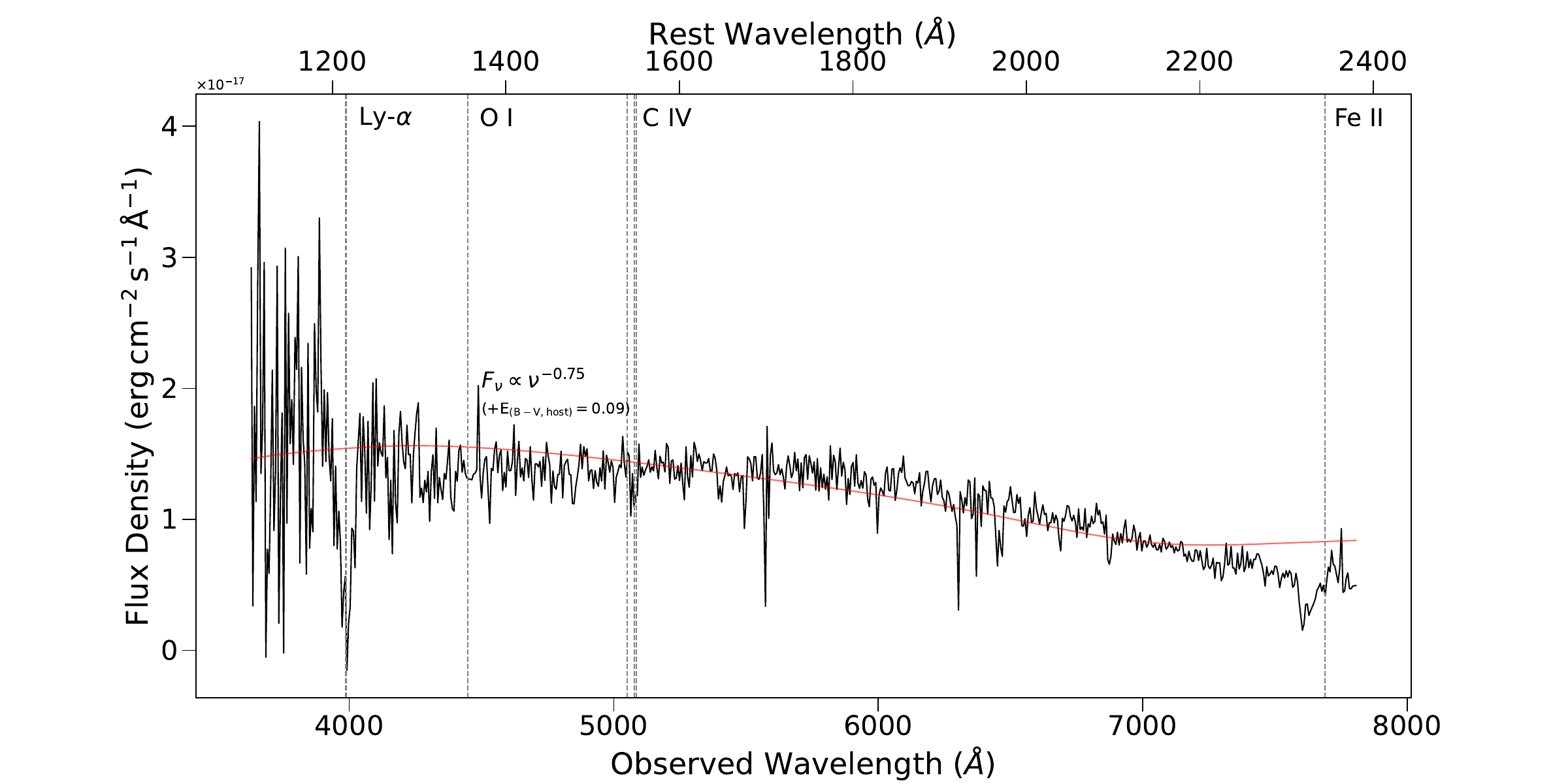}
    \caption{Optical spectrum of AT 2023sva, obtained by the OSIRIS instrument (details in \S \ref{SEDoptical}). We show the Lyman-$\alpha$ abosorption feature along with absorption features characteristic to GRB afterglows at a redshift $z = 2.280 \pm 0.002$, and a best-fit spectral power-law model with additional host-galaxy extinction (details in \S \ref{SEDoptical}). }
    \label{spectrum}
\end{figure*}

\subsection{Search for Associated GRB}
\label{grbsearch}
The $\gamma$-ray sky is monitored by the third Interplanetary Network (IPN), whose most sensitive instruments are the \textit{Swift} Burst Alert Telescope (BAT), \textit{Fermi} Gamma-ray Burst Monitor (GBM; \citealt{Meegan+2009}), and the Konus-Wind \citep{Barthelmy2005} instrument. We searched the archives of these three instruments, as well as the  AstroSat Cadmium Zinc Telluride Imager (CZTI; \citealt{Bhalerao2017}) instrument, to determine if there is an associated GRB temporally and spatially coincident with AT 2023sva. We searched the time periods between the last ZTF non-detection and the first detection (see \S \ref{ZTFdisc}). Though there were several GRBs in this time period, we found no \textit{Fermi} or \textit{Swift} GRBs temporally and spatially coincident to AT 2023sva, through searching the \textit{Fermi} GBM Burst Catalog, the \textit{Fermi} GBM Subthreshold Trigger list, and the \textit{Swift} GRB Archive. There were reports of a candidate GRB on the Gamma-Ray Coordinates Network archives; however this counterpart was later classified as a likely solar flare and deemed unrelated to AT 2023sva \citep{GCNsolar}. Due to the 2 day time period between the last non-detection and first detection, there was a significant amount of time that both \textit{Fermi} and \textit{Swift} were not viewing the field of AT 2023sva due to being occulted by the Earth. 
%Becuase there is a 2 day time period between the last non-detection and first detection, the position of AT 2023sva was visible to both satellites for ample time during these time periods.

However, because of Konus-Wind's 4$\pi$ FOV and its interplanetary orbit at the Earth-Sun L1 Lagrange point, AT 2023sva's location was always visible to the instrument. Konus-Wind detected two GRBs not detected by \textit{Fermi} during the time period of interest. The first, which occurred on 2023-09-15 06:54:20, resulted in Konus-Wind's ecliptic latitude response clearly being inconsistent with AT 2023sva's position. Both Konus-Wind and AstroSat-CZTI detected the second GRB, GRB 230917A on 2023-09-17 00:44:38.873 by Konus-Wind and on 2023-09-17 00:44:43.5 by Astrosat-CZTI \citep{Navaneeth2023}. We utilized the propagation time delay between Konus-Wind and AstroSat-CZTI and the Konus-Wind ecliptic latitude response to calculate the 3$\sigma$ IPN localization region and determine if it is consistent with the position of AT 2023sva.

We show the localization region in Figure \ref{IPNlocalization}, along with contours showing various Earth-occulted regions for the different satellites, along with the $>10 \%$ coded FOV of \textit{Swift}. We see that GRB 230917A's 3$\sigma$ localization region is clearly not consistent with AT 2023sva's location. The lack of detections from \textit{Fermi} and \textit{Swift} are also consistent with this localization, as GRB 230917A's location was occulted by the Earth for \textit{Fermi} and was outside \textit{Swift}'s coded FOV. We also localized the burst through the AstroSat-CZTI localization framework \citep{Saraogi2024}. Since the burst was not bright, the localization was quite coarse, placing AT 2023sva within the 57\% contour. This also supports the possibility that the burst may not be associated with the optical transient. Therefore, we determine that AT 2023sva does not possess a detected GRB counterpart. 

%On the other hand, AT 2023sva's position is not occulted by the Earth for \textit{Fermi}, so if a bright burst was associated with the transient, we would have expected \textit{Fermi} to detect it. 
We calculate the AstroSat-CZTI upper limits in the time window between the last non-detection and first detection (a window size of ~21600 s; with a confidence of 97.44\%), utilizing the methodology in \citet{Sharma2021}. We analyzed AstroSat-CZTI's duty cycle during the 50+ hour window ($\sim$ 19 AstroSat-CZTI orbits) between the first non-detection and detection. With the source's high declination, AstroSat had continuous visibility of the source, but $\sim$ 29\% of the time was lost to South Atlantic Anomaly downtime and 3\% to slewing.
AstroSat-CZTI detects a confident GRB approximately every 120 hours. Our extended time window significantly increases the false alarm rate, leading to the flux upper limits having a relatively low confidence ($\sim$ 87\%). Additionally, the method for estimating flux limits requires 10 ``witness” neighboring orbits, which is insufficient to reliably estimate the background for our time window. As a result, the upper limits we derived were not meaningful.

We then utilize the Konus-Wind
non-detection to derive a peak flux upper limit for AT 2023sva’s GRB counterpart. During the interval of interest Konus-Wind was continuously observing the whole sky in the waiting mode with temporal resolution 2.944 s. In this mode count rates are recorded in the three energy bands: 19--80 keV (G1),
80--325 keV (G2), and 325--1290 keV (G3). The instrument background count
rate varied slowly at timescale of a day at  $< 7
\%$. There were a number of minor data gaps of a size of about a few $\times$ 2.944 s, accounting for 4 \% of the time between the last non-detection and first detection.
After the trigger on GRB 230917A, there was an hour long data gap in the waiting mode record due to instrument data readout. During this interval, only the G2 count rate was available with $\sim$ 4 s resolution.

Using the waiting mode data free from GRBs detected in the time period between the last non-detection and first detection of AT 2023sva and the response for AT 2023sva's position, we estimate an upper limit ($90\%$ confidence) on the 20--1500
keV peak flux to $1.5\times10{^{-7}} \, \rm{erg \, cm^{-2} \, s^{-1}}$ for a typical LGRB spectrum
(a Band function with $\alpha =-1$, $\beta =-2.5$, $E_{\rm{peak}}$ =300 keV) and a 2.944 s timescale. The lack of a detected GRB counterpart suggests that AT 2023sva's limiting fluence is comparable to that of the weakest
burst from \citet{Tsvetkova2021}, who did a study on Konus-Wind bursts simultaneously detected by \textit{Swift}-BAT. This fluence is $4 \times 10^{-7} \rm{erg \, cm^{-2}}$ and given the redshift $z = 2.28$, this corresponds to an isotropic equivalent energy upper limit of $E_{\gamma, \rm{iso}} < 1.6 \times 10^{52} $ erg for AT 2023sva's GRB counterpart. 
%however Konus-Wind does not provide localizations, and there is no localization for this event provided by \textit{Swift} or \textit{Fermi}. Given the rate of Konus-Wind GRB detections (0.42 d$^{-1}$), the observation of 1 GRB within a 2 day search window is consistent with random fluctuations in Konus-Wind data. Therefore, we determine that AT 2023sva does not possess a detected GRB counterpart. Following the analysis of \citet{Cenko2013}, we take the nominal fluence threshold of the IPN between 50 and 300 keV as $10^{-6}$ erg cm${^{-2}}$. Given the redshift $z = 2.28$, using the luminosity distance, this corresponds to an isotropic equivalent energy upper limit of $E_{\gamma, \rm{iso}} < 4 \times 10^{52} $ erg.
\section{Optical and Radio Analysis}
\label{opradioanalysis}

\begin{figure*}
    \centering
    \includegraphics[width=0.75\linewidth]{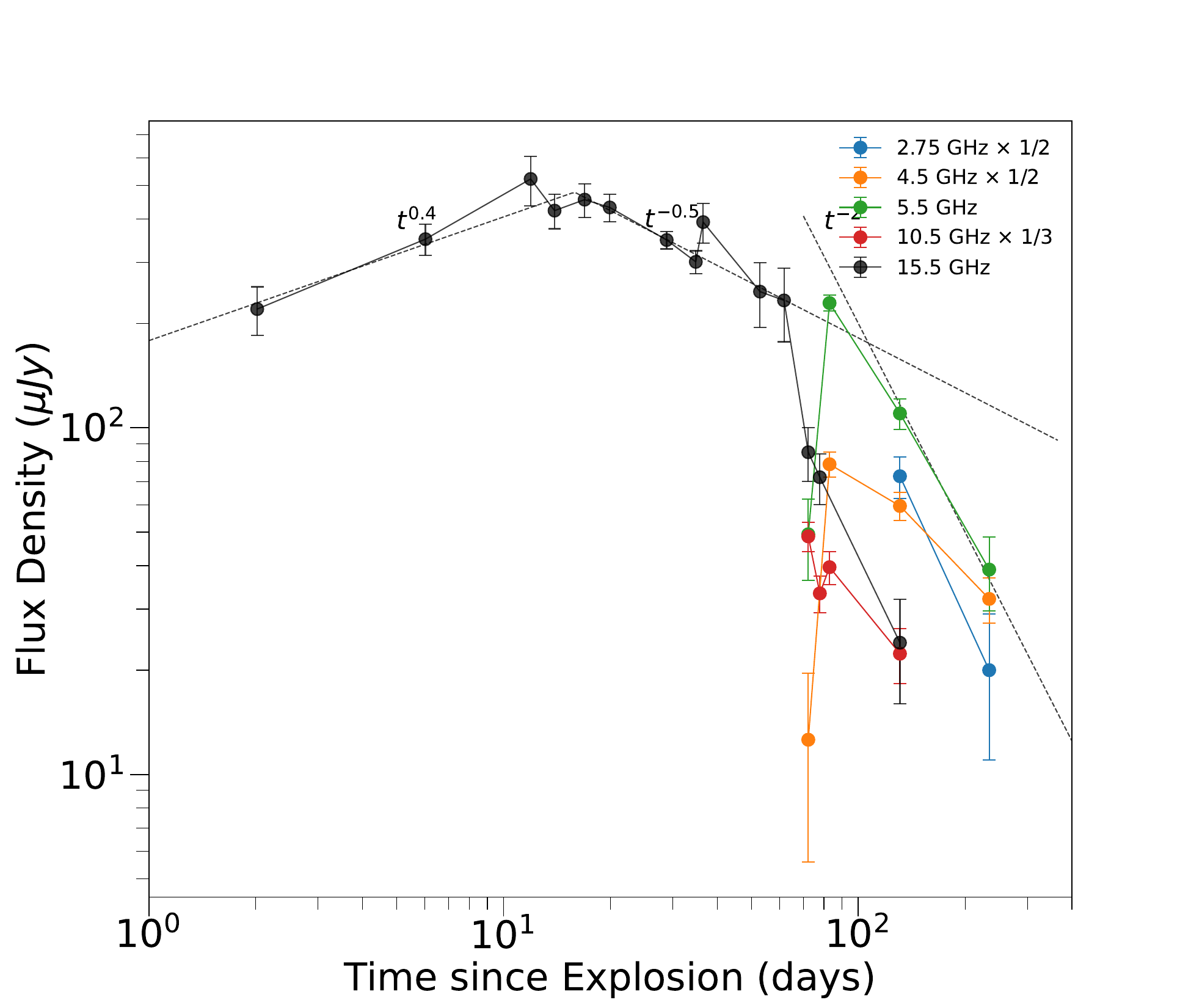}
    \caption{The AMI (15.5 GHz) and VLA (15.5 GHz and other frequencies) radio LCs with respect to the observed time since explosion. The 15.5 GHz LC is the most well-sampled over time and we fit a power-law to the initial rise, shallow decay, and then the steep decay. This behavior is most likely attributed to the peak synchrotron frequency passing through the radio bands and then a jet break at late times. The lower frequency radio LCs show high variability, due to the presence of interstellar scintillation.}
    \label{radioLC}
\end{figure*}

\subsection{Optical LC}
\label{opticalLCsection}
We show the optical LC in Figure \ref{opticalLC} and the optical transient fades very rapidly. There is a possible break in the LC that occurs between one to two days after explosion; however, this break cannot be constrained because of the transient's rapidly fading nature. This leads to a lack of detections at later times to truly constrain the presence of a break (or lack thereof). Therefore, we fit a simple power-law to the early-time data in $g$, $r$, and $i$ bands simultaneously with different normalization factors, $F_\nu \propto (t-t_0)^{-\alpha}$, where $\alpha$ is the power-law decay index, and $t_0$ is the estimated time of explosion. We set $t_0$ equal to the best-fit time of explosion derived in \S \ref{Modeling}, of $t_0 = 60204.09$ MJD, or 7.5 hours prior to the first detection. We refer to this $t_0$ as the time of explosion for the rest of this work. We note that we did try to let $t_0$ be a free parameter in the fit as well. Through this fitting procedure, we found that $t_0$ was constrained to the midpoint between the last non-detection and the first detection, likely due to the small number of data points. Before fitting a power-law, we first correct the optical magnitudes for both the Milky Way ($E(\rm{B-V})_{\rm{MW}} = 0.24$ mag), and host galaxy ($E(B-V)_{\rm{host}} = 0.09$; see \S \ref{SEDoptical}) extinction. We derive a power-law decay index $\alpha = 1.64 \pm 0.02$. 

From our best-fit LC, we see that the single power-law decay model does not fit the last epoch of observations well. This is due to the possible break in the LC mentioned earlier, which is likely from the synchrotron cooling frequency $\nu_{\rm{c}}$ passing through the optical bands, resulting in a change of the temporal power-law decay index of the LC \citep{Granot2002, Ryan2020}. We provide more evidence for this in \S \ref{Modeling}.

%This is likely due to one $r$ and one $g$ band point showing an increase in flux with respect to the previous epoch of observations (see Figure \ref{opticalLC}). This increase in flux is not normally seen in the standard picture of optical afterglows, and is most likely due to flaring activity \citep{Yi2017}.

\subsection{Optical SED and Spectroscopy}
\label{SEDoptical}

\begin{figure*}
    \centering
    \includegraphics[width=0.99\linewidth]{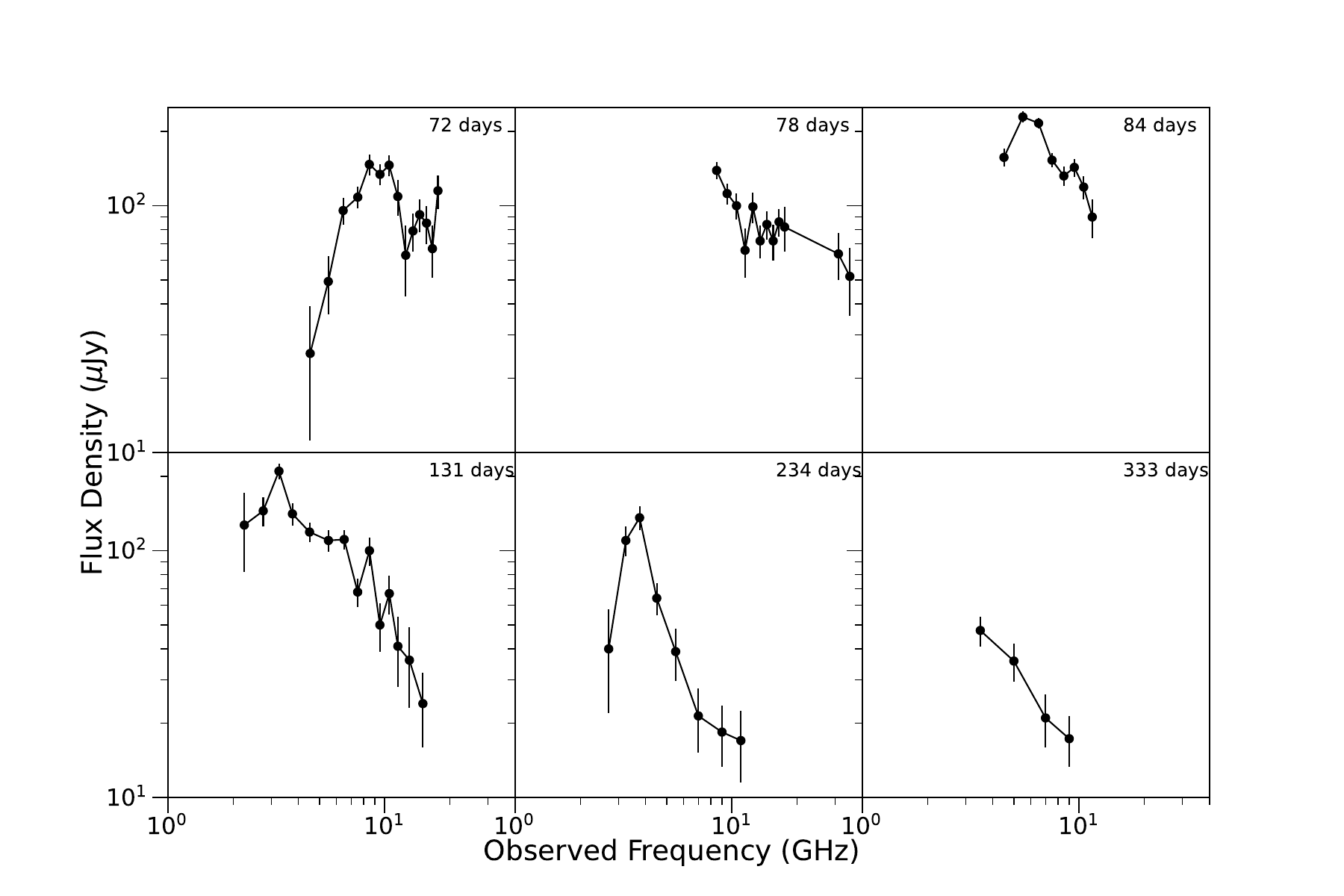}
    \caption{VLA Radio SEDs of AT 2023sva, with the observer frame time of the epochs with respect to the best-fit explosion time shown in the plots. The spectrum at 72 days shows multiple sharp spectral breaks due to the presence of interstellar scintillation. The sharp breaks continue at lower frequencies until the epoch at 131 days.}
    \label{radiospectra}
\end{figure*}
We fit the $gri$ data from SEDM taken over three separate epochs (detailed in Table \ref{photometrytable}) to a spectral power-law model $F_\nu \propto \nu^{-\beta}$, after correcting for the Milky Way extinction. We obtain a spectral index of $\beta = 0.95 \pm 0.15$. When including the X-ray upper-limit in the fitting extrapolated to the midpoint of the SEDM observations, we constrain the optical to X-ray spectral index $\beta_{\rm{OX}} > 1.1$. Therefore, we determine that the X-ray and optical data likely do not lie on the same spectral segment. 

In Figure \ref{spectrum}, we show the OSIRIS spectrum of AT 2023sva (details of observations in \S \ref{spectroscopy}). The spectrum shows a clear Lyman-$\alpha$ feature and various other absorption features, at a redshift $z = 2.280 \pm 0.002$. We measure the equivalent widths (EWs) of the \ion{O}{I}, \ion{C}{IV}, and \ion{Fe}{II} absorption features. We find strengths of $1.9 \pm 1.1$, $2.63 \pm 0.70$, and $1.42 \pm 0.56$ $\AA$. We then compare the line strengths of AT 2023sva with other GRBs in the literature through calculating the line strength parameter (LSP), described in \citet{Ugarte2012}. We calculate a LSP of $-1.70 \pm 0.58$, which corresponds to the 1.03th percentile of GRBs. Therefore, 99\% of the GRBs in the literature have line strengths greater than AT 2023sva, pointing towards an extremely low-density sight line to the source.

Because only an X-ray upper limit was obtained, we could not make any inferences about the presence of host-galaxy extinction from the SED fitting. However, in Figure \ref{spectrum}, we see that the Milky Way extinction-corrected spectrum shows a distinct curvature that is likely due to extra extinction from the host galaxy. Therefore, we fit a power-law to the optical spectrum with the addition of host-galaxy extinction as a free parameter, using the \citet{ccm1989} extinction law, with $R_v = 3.1$. We find a best-fit spectral index of $\beta = 0.75 \pm 0.07$ and $E(B-V)_{\rm{host}} = 0.09 \pm 0.01$ mag. We note that there is no prominent 2175 \r{A} feature in the spectrum, which is a characteristic feature of the \citet{ccm1989} model. It has also been shown that GRB host galaxies in general rarely show strong evidence for this feature, though in most cases it also cannot be ruled out \citep{Schady2012}. Therefore, the uncertainties reported here are statistical uncertainties and there are likely larger uncertainties dominated by the use of a particular extinction model. 

\subsection{Closure Relations }
\label{closure}
There are characteristic closure relations between $\alpha$, $\beta$, and $p$ (the electron spectral index) that correspond to different astrophysical environments (a constant density ISM environment or a stellar wind environment), as well as cooling regimes within the synchrotron spectrum \citep{Sari1998, Granot2002}. We test the values derived for $\alpha$ and $\beta$ within these regimes, assuming a standard, tophat jet structure. First, we determine whether we are in the fast or slow cooling regime. In the fast cooling regime, the synchrotron frequency corresponding to the minimum Lorentz factor that electrons are accelerated to in the shockwave in the context of a power-law distribution ($\nu_{\rm{m}}$, also known as the peak frequency) is larger than the synchrotron cooling frequency $\nu_{\rm{c}}$, so most electrons are expected to quickly cool to $\nu_{\rm{c}}$. In this regime, the optical bands can either be below $\nu_{\rm{m}}$ ($\nu_{\rm{c}} < \nu < \nu_{\rm{m}}$) or above $\nu_{\rm{m}}$ ($\nu_{\rm{c}} < \nu_{\rm{m}} < \nu$). For the case where $\nu_{\rm{c}} < \nu < \nu_{\rm{m}}$, $\beta = 0.5$ \citep{Sari1998, Granot2002} for both the ISM and wind environment. For the case where $\nu_{\rm{c}} < \nu_{\rm{m}} < \nu$, $\beta = p/2$ for both environments \citep{Sari1998, Granot2002}. It is clear that the $\beta$ we derive is not consistent with the $\nu_{\rm{c}} < \nu < \nu_{\rm{m}}$ case. Furthermore, if the optical bands were above the peak frequency, then that would imply $p = 1.5 \pm 0.14$. This is an abnormally small, non-physical value for $p$, which is generally expected to be between 2 and 3 \citep{Curran2010}. Therefore, this implies that we are not in a fast cooling environment.

For a slow cooling environment, $\nu_{\rm{m}}$ is less than $\nu_{\rm{c}}$ and electron cooling is not efficient. In this regime the optical bands can be either below or above $\nu_{\rm{c}}$. If $\nu_{\rm{m}} < \nu < \nu_{\rm{c}}$, then $\beta = (p-1)/2$, for both an ISM and wind environment. This would imply $p = 2.50 \pm 0.14$, which is a reasonable value. If $\nu_{\rm{m}} < \nu_{\rm{c}} < \nu$, $\beta = p/2$ for both environments, this implies $p = 1.50 \pm 0.14$, which again is an unreasonable value. So from this analysis, we determine that assuming a tophat jet, we are in a slow cooling environment where the optical bands are below the cooling frequency. For a constant density ISM environment, $\alpha = 3\beta/2$ in this regime \citep{Sari1999}, which would imply $\alpha = 1.13 \pm 0.15$, which clearly does not match the derived value of $\alpha = 1.64 \pm 0.02$. For the wind environment, $\alpha = (3\beta +1 )/2$ in this regime, which would imply $\alpha = 1.63 \pm 0.15$. This is consistent with our observed $\alpha$. However, the jet's structure may be more complex than just a tophat and we revisit the closure relations within this context in \S \ref{Modeling}.

\subsection{Radio LC and SED}
\begin{table}
    \centering
\caption{AMI radio observations of AT 2023sva, obtained at 15.5 GHz.}
    \begin{tabular}{l|l|l}
    \hline
    \hline
        MJD & $F_\nu$ ($\mu \rm{Jy}) $ & $\sigma$ ($\mu \rm{Jy}$) \\ 
        \hline
        60206.11 & 220 & 41 \\ 
        60210.11 & 350 & 37 \\ 
        60216.01  & 522 & 106 \\ 
        60218.01 & 423 & 55 \\ 
        60221.01 & 455 & 52 \\ 
        60224.01 & 432 & 80 \\ 
        60232.91 & 348 & 33 \\ 
        60238.91 & 301 & 47 \\ 
        60240.61 & 392 & 90 \\ 
    \end{tabular}
    \label{AMI}
\end{table}

\begin{table}
\centering
\caption{VLA observations of AT 2023sva.}
    \begin{tabular}{l|l|l}
    \hline
    \hline
    $\nu$ (GHz) & $F_\nu$ ($\mu \rm{Jy}) $ & $\sigma$ ($\mu \rm{Jy}$) \\
    \hline
        Epoch 1 (MJD 60276.35) & & \\
        \hline
        4.53 & 25.2 & 14 \\ 
        5.49 & 49.3 & 13 \\ 
        6.43 & 95.6 & 12 \\ 
        7.51 & 108.3 & 11 \\ 
        8.49 & 147 & 14 \\ 
        9.51 & 134 & 13 \\ 
        10.48 & 146 & 14 \\ 
        11.49 & 109 & 18 \\ 
        12.5 & 63 & 20 \\ 
        13.5 & 79 & 14 \\ 
        14.5 & 92 & 14 \\ 
        15.6 & 85 & 15 \\ 
        16.6 & 67 & 16 \\ 
        17.6 & 115 & 18 \\ 
        \hline
        Epoch 2 (MJD 60282.00) & ~ & ~ \\
        \hline
        8.5 & 139 & 11 \\ 
        9.5 & 112 & 11 \\ 
        10.5 & 100 & 12 \\ 
        11.5 & 66 & 15 \\ 
        12.5 & 99 & 14 \\ 
        13.5 & 72 & 11 \\ 
        14.5 & 84 & 11 \\ 
        15.5 & 72 & 12 \\ 
        16.5 & 86 & 11 \\ 
        17.5 & 82 & 17 \\ 
        31 & 63.8 & 13.9 \\ 
        35 & 51.7 & 16 \\ 
        \hline
        Epoch 3 (MJD 60287.06) & ~ & ~ \\ 
        \hline
        4.5 & 157 & 13 \\ 
        5.5 & 229 & 12 \\ 
        6.5 & 216 & 11 \\ 
        7.5 & 153 & 10 \\ 
        8.5 & 132 & 12 \\ 
        9.5 & 143 & 12 \\ 
        10.5 & 119 & 13  \\ 
        11.5 &90 & 16  \\
        \hline
        Epoch 4 (MJD 60335.10) & ~ & ~ \\ 
        \hline
        2.25 & 127 & 45 \\ 
        2.75 & 145 & 20 \\ 
        3.25 & 210 & 15 \\ 
        3.75 & 141 & 15 \\ 
        4.5 & 119 & 11 \\ 
        5.5 & 110 & 11 \\ 
        6.5 & 111 & 10 \\ 
        7.5 & 68 & 9 \\ 
        8.5 & 100 & 13 \\ 
        9.5 & 50 & 11 \\ 
        10.5 & 67 & 12 \\ 
        11.5 & 41 & 13 \\ 
        13 & 36 & 13 \\ 
        15 & 24 & 8 \\ 
        \hline
        Epoch 5 (MJD 60437.90) & ~ & ~ \\
        \hline
        2.7 & 40 & 18 \\ 
        3.24 & 110 & 15 \\ 
        3.75 & 136 & 15 \\ 
        4.5 & 64.2 & 9.6 \\ 
        5.5 & 39 & 9.4 \\ 
    \end{tabular}
    \label{VLA}
\end{table}

\begin{table}
\centering
\begin{tabular}{l|l|l}
        7 & 21.4 & 6.2 \\ 
        9 & 18.4 & 5.1 \\ 
        11 & 17 & 5.5 \\
        \hline
         Epoch 6 (MJD 60537.30) & ~ & ~ \\ 
         \hline
        3.5 & 47.5 & 6.6 \\ 
        5 & 35.7 & 6.2 \\ 
        7 & 21 & 5.1 \\ 
        9 & 17.3 & 4 \\ 
 \end{tabular}
\end{table}

\begin{figure}
    \centering
\includegraphics[width=0.99\linewidth]{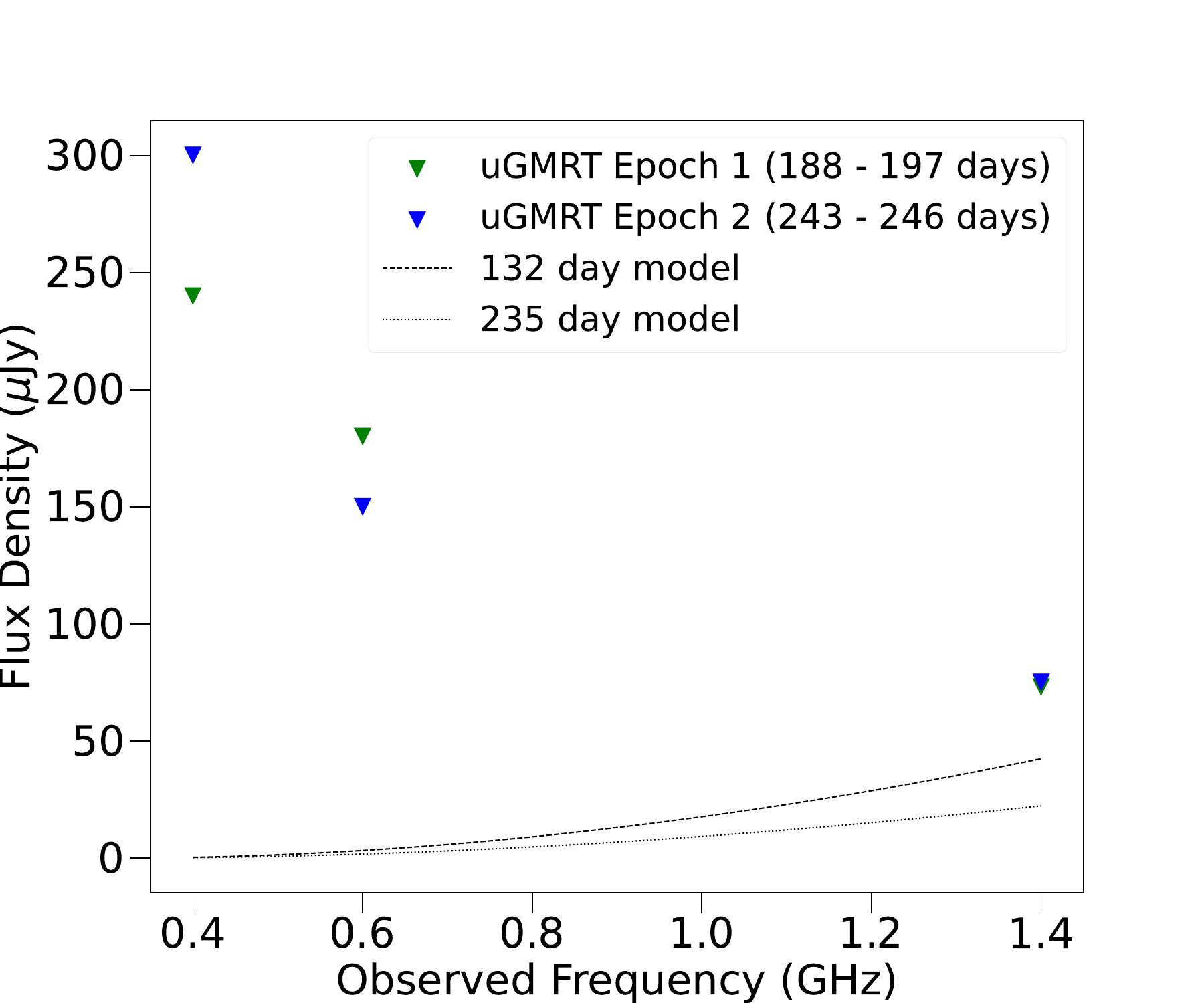}
    \caption{uGMRT observations of AT 2023sva, where the upside down markers represent the $3\sigma$ derived upper limits. Two power-law models corresponding to the low-frequency VLA observations at 131 and 234 days are also shown and the upper limits are consistent with the VLA observations. }
    \label{ugmrt}
\end{figure}

The radio LC at select frequencies is shown in Figure \ref{radioLC}. We see that the 15.5 GHz LC shows a gradual rise ($t^{0.4}$) at early times, until around ten days after the explosion. The LC then exhibits a shallow decay ($t^{-0.5}$) and thereafter transitions to a steep decay ($t^{-2}$), seen at lower frequencies as well. Lower frequency ($<$ 11 GHz) observations begin at around day 70 and there is significant short-term variability in the first couple of epochs in the 4 and 5 GHz LCs, likely indicative of strong interstellar scintillation (ISS). The initial rise in the 15.5 GHz LC until around 10 days and then shallow decay until around 70 days, can be attributed to the spectral break $\nu_{\rm{m}}$ passing through the 15.5 GHz band. The subsequent steepening across all frequencies is most likely attributed to a jet break. We model the radio LC in \S \ref{Modeling}.

In Figure \ref{radiospectra}, we show the coeval radio SEDs, at six different epochs presented in the observer frame. We see that the SED at 72 days has multiple sharp spectral breaks. This is due to ISS and the modulations are a factor of $\sim$ 2. We see this behavior continues to persist at lower frequencies until the epoch at 131 days, though the strength of the scintillation diminishes.

%continues to exist in the SED taken 234.5 days after the explosion at the lowest frequencies, and the only spectrum free from modulations is the SED at 333.9 days, though there is only limited coverage with 4 points. 

We expect $\nu_{\rm{m}}$ to pass through the radio bands over time, leading to spectral breaks in the SEDs. Due to the variability caused by ISS, we were not able to identify any clear breaks in the first two epochs. However, starting from the epoch at 84 days, we see that the spectrum shows a clear rise and peak at low frequencies and the location of the peak moves lower in frequency space, until it is not visible anymore in the 333 day spectrum, starting from $\sim$ 10 GHz in the 72 day epoch to $\sim$ 3 GHz in the 234 day epoch. We likely are still below the cooling frequency $\nu_{\rm{c}}$ at the times the SEDs were taken, as we do not see a steepening of the power-law at later times.  We fit a spectral power-law model ($F_\nu \propto \nu^{-\beta}$) to every spectral epoch except for the 72 day epoch, both to the regions below and above $\nu_{\rm{m}}$. For frequencies below $\nu_{\rm{m}}$, we find an average value of $\beta = 2.2$, with a range of $1.4$ to $4.1$, and for frequencies above $\nu_{\rm{m}}$, we find an average value of $\beta = -1.2$, with a range of $-0.5$ to $-2.1$. We note that the spectral indices derived below $\nu_{\rm{m}}$ have large error bars and are poorly constrained, as there are only two points below the peak frequency in the 84 day epoch and three points in the 131 day and 234 day epochs.

In Figure \ref{ugmrt}, we also show the low frequency uGMRT observations. Radio emission was not detected at the source position in any of the GMRT maps providing 3-sigma flux density limits of $<75\,\mu$Jy at 1.4 GHz, $<180\,\mu$Jy at 0.6 GHz, and $<300\,\mu$Jy at 0.4 GHz at the source position. The first observation took place between 188 and 197 days after $t_0$ and the second took place between 243 and 246 days after $t_0$. The upper limits derived are consistent with the power-laws derived from fitting the 131 day and 235 day epochs.

\begin{figure}
    \centering
    \includegraphics[width=0.99\linewidth]{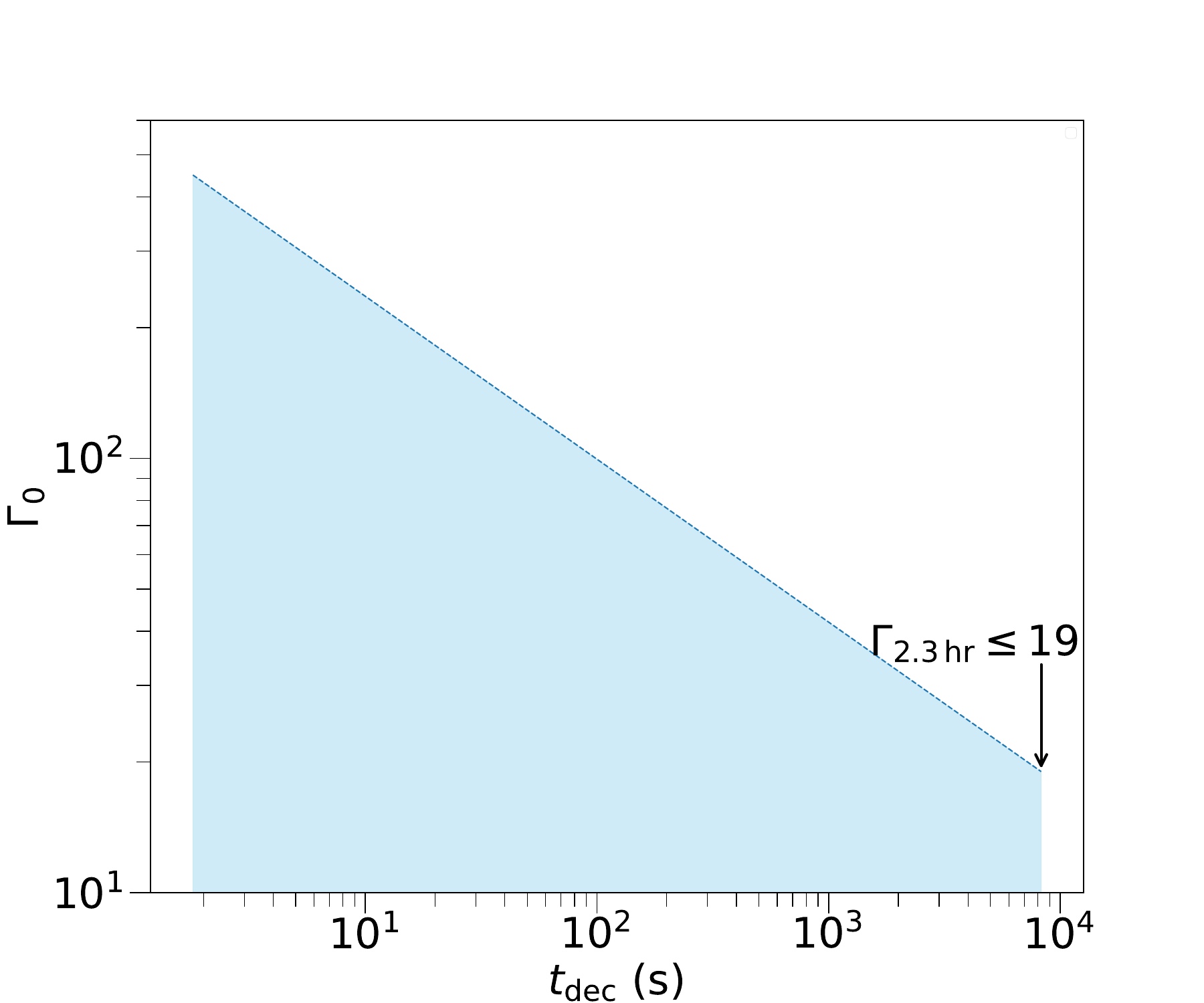}
    \caption{Upper limit on the initial Lorentz factor $\Gamma_0$ plotted against the assumed rest-frame deceleration time $t_{\rm{dec}}$. The limit is derived from the limit obtained from scintillation analysis in \S \ref{scintillation} and assuming $\Gamma \propto t^{-3/8}$. The allowed values are shown as the shaded region.}
    \label{gamma_in}
\end{figure}

\begin{figure}
    \centering
    \includegraphics[width=0.99\linewidth]{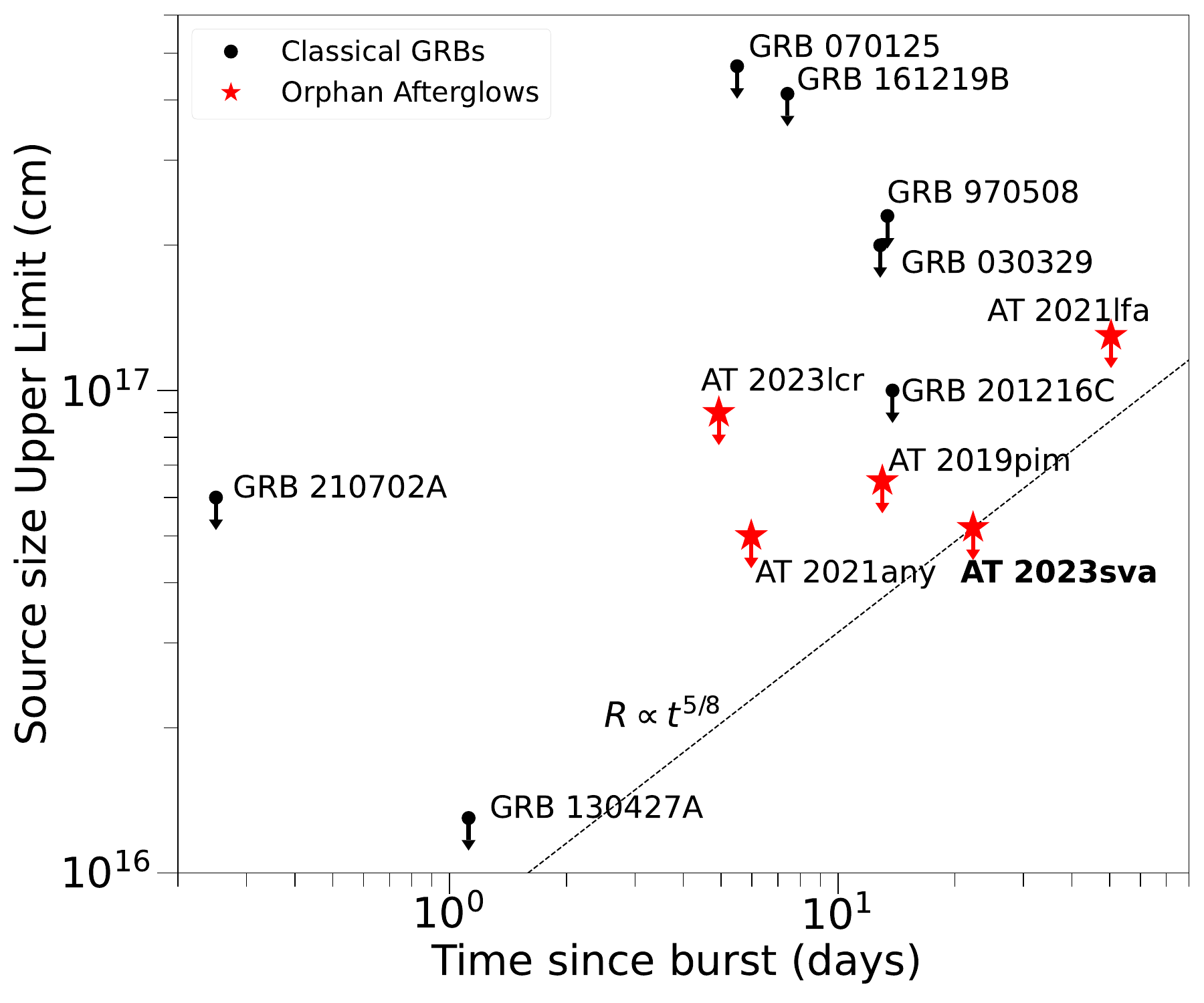}
    \caption{Comparison of source size upper limits derived from ISS analyses in seven classical GRBs (GRB 907508, GRB 030329, GRB 070125, GRB 130427A, GRB 161219B, GRB 210702A, and GRB 201216C), and five orphan afterglows (AT 2019pim, AT 2021any, AT 2021lfa, AT 2023lcr, and AT 2023sva), with respect to their rest-frame times since explosion. We also show AT 2023sva's source size upper limit as a function of time, assuming a constant density ISM environment.}
    \label{sourcesize}
\end{figure}

%However, all of the SEDs with the exception of the 72.9 day epoch show a clear decline in flux with respect to frequency, suggesting that $\nu_{\rm{m}}$ has already passed through the radio bands. It is possible the first epoch at 72.9 days likely also exhibits this behavior, though the modulations at lower frequencies do not make it clear.

\begin{figure*}
    \centering
    \includegraphics[width=0.49\linewidth]{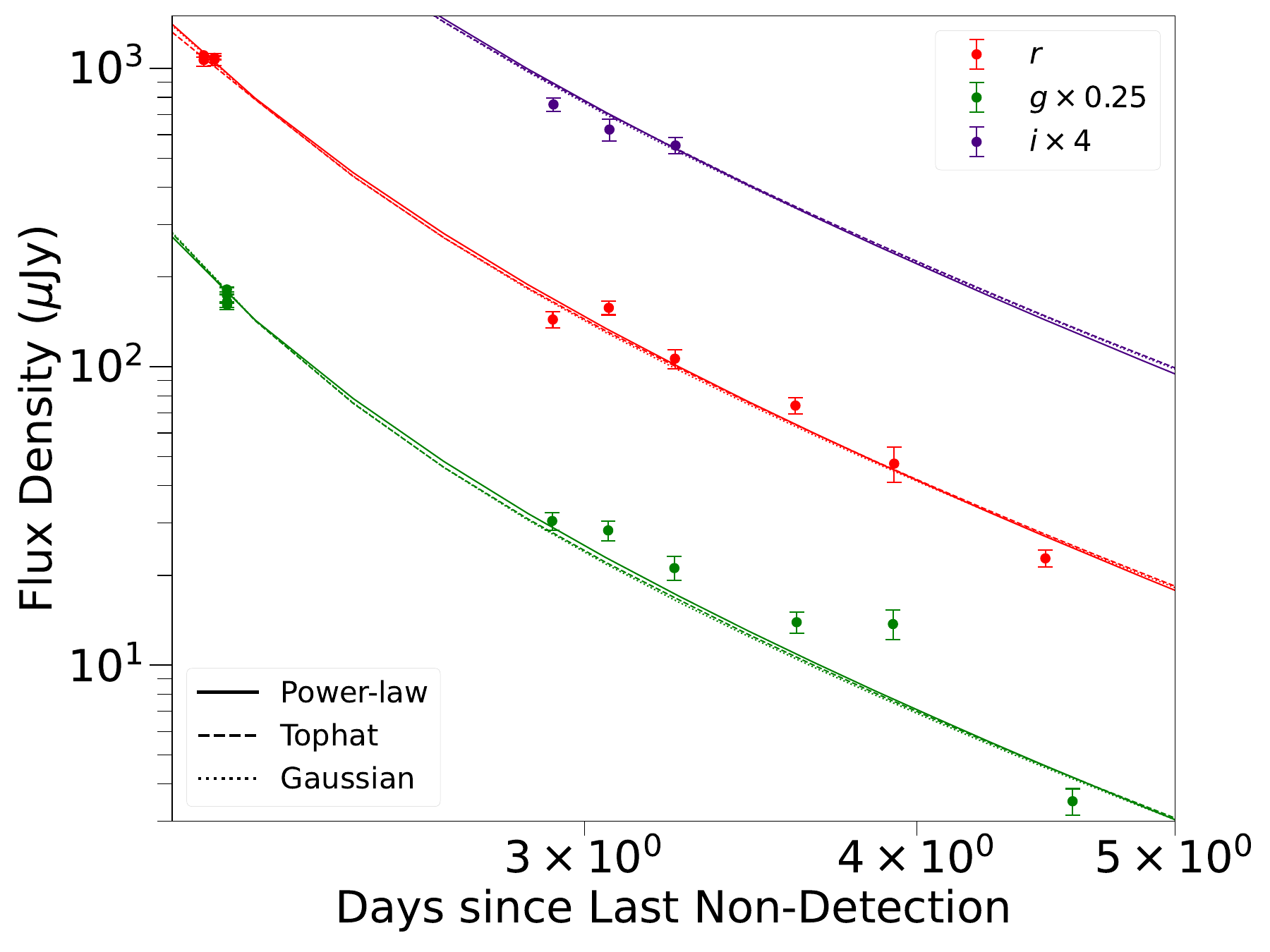}
    \includegraphics[width=0.49\linewidth]{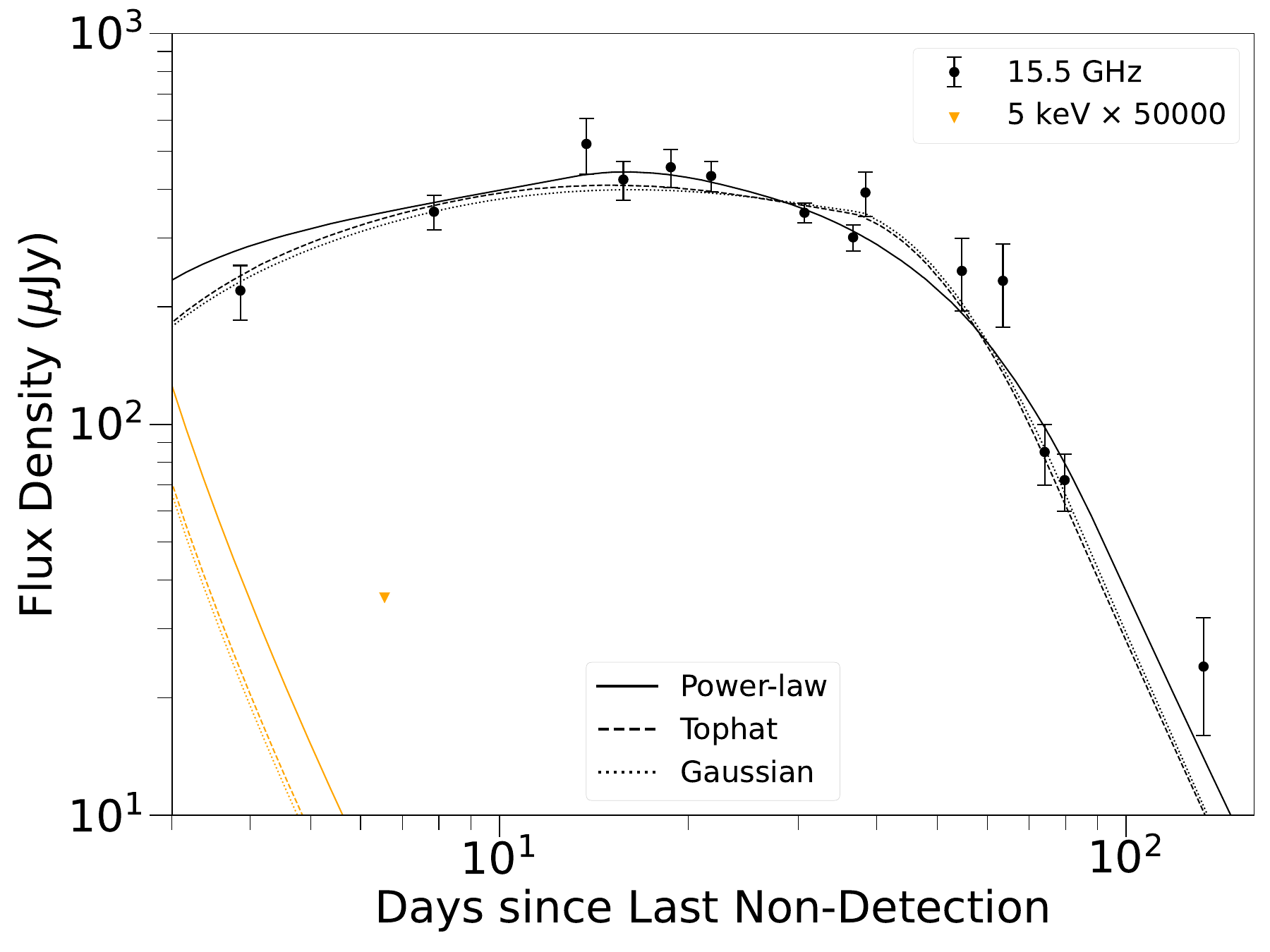}
    \caption{\textit{Left panel}: The optical observations of AT 2023sva, along with the best-fit power-law structured jet, tophat jet, and Gaussian structured jet fit to the observed LC. Optical fluxes have been multiplied by a constant factor for viewing purposes and have been corrected for MW and host galaxy extinction. \textit{Right panel}: The 15.5 GHz radio observations of AT 2023sva and the X-ray upper limit, along with the best-fit power-law structured jet, tophat jet, and Gaussian structured jet fit to the observed LC. The X-ray limit has been multiplied by a constant factor for viewing purposes. }
    \label{modeling}
\end{figure*}

\subsection{Constraints from Radio Scintillation}
\label{scintillation}
The radio SED shows clear variations in frequency space in the earliest epoch at 72 days, across all frequencies. The radio LC also shows clear modulations by a factor of $\sim$ 2 at lower frequencies up to 100 days. We interpret these variations as being due to ISS, which is the result of small-scale inhomogeneities in the ISM, which change the phases of incoming wavefronts of radio sources. Because the line of sight from Earth to the source changes as the Earth moves, this causes a change in flux, due to scattering from electrons along the line of sight through our Galaxy. There is a characteristic transition frequency $\nu_{\rm{ss}}$ where strong scattering transitions ($\nu < \nu_{\rm{ss}}$) to weak scattering ($\nu > \nu_{\rm{ss}}$). GRB radio observations are expected to be affected by ISS (e.g., \citealt{Granot2014}) and analyzing their ISS can provide insights towards their angular size and Lorentz factors.

We utilize a similar method that \citet{Perley2024} used for AT 2019pim, to derive AT 2023sva's source size. From Figures 1-2 of \citet{Walker1998} (erratum \citealt{Walker2001}), we determine that the transition frequency $\nu_{\rm{ss}}$ along the line of sight of AT 2023sva is $\sim$ 15 GHz and the Fresnel scale at that frequency is $\theta_{F0} \approx 2 \, \rm{\mu arcsec}$. At the angular diameter distance of AT 2023sva, this corresponds to a physical size of $5.2 \times 10^{16}$ cm. From the radio SED in Figure \ref{radiospectra}, we see that there is strong ISS at  $\nu_{\rm{ss}}$ during the earlierst epoch at 72 days and it significantly decreases in later epochs. For strong ISS to exist near $\nu_{\rm{ss}}$, the source size must be comparable or smaller than the physical size corresponding to the Fresnel scale. Therefore, we estimate AT 2023sva's physical size to be at most $5.2 \times 10^{16}$ cm at 72 days. Converting this to an average Lorentz factor in the rest-frame, we get $\Gamma_{\rm{av, 22.0 \, d}} \leq 2.4$. In the ISM, as the jet is decelerating, the projected size increases over time as $R \propto t^{5/8}$, or $\Gamma \propto t^{-3/8}$ \citep{Galama2003}. Extrapolating the ISS limit to the time of first detection (7.5 hours after explosion in the observer frame, or 2.3 hours after in the rest-frame), we derive a Lorentz factor upper limit of $\Gamma_{\rm{av, 2.3 \, hr}} \leq 19$.

However, $\Gamma$ at the time of first detection is different than the initial Lorentz factor, $\Gamma_0$, as it is dependent on the time of explosion and the initial deceleration timescale, $t_{\rm{dec}}$. If we assume that $t_{\rm{dec}} \sim$ 20 s \citep{Ghirlanda2022} from the explosion epoch (corresponding to the typical rest-frame timescale for a GRB optical LC to peak) and also assume that the Lorentz factor is constant from the beginning of the explosion up to the beginning of deceleration (coasting phase), we derive an initial Lorentz factor limit at the start of deceleration of $\Gamma_0 < 178$. We note that this assumption has many caveats, as it has been shown that $t_{\rm{dec}}$ varies greatly for GRBs \citep{Ghirlanda2018}. Therefore, in Figure \ref{gamma_in} we show the parameter space for allowed $\Gamma_0$, given our scintillation limit, for a range of $t_{\rm{dec}}$ starting from the last non-detection and ending with the first detection. We see that if $t_{\rm{dec}} \gtrsim 94$ s, that $\Gamma \lesssim 100$, which would indicate a more moderately relativistic outflow than the known population of cosmological GRBs. However, our observations are not constraining enough to make this claim.

There have only been a handful of scintillation  measurements for GRBs confirmed in the literature, as they necessitate high-cadence, multi-frequency radio observations of GRBs out to late times. There have been seven classical GRBs whose source sizes were able to be constrained by the presence of strong ISS (GRB 970508, \citealt{Frail2000}; GRB 030329, \citealt{Berger2003,Taylor2004, Taylor2005, Philstrom2007}; GRB 070125, \citealt{Chandra2008}; GRB 130427A, \citealt{vanderhorst2014}; GRB 161219B, \citealt{Alexander2019}; GRB 210702A, \citealt{Anderson2023}; and GRB 201216C, \citealt{Rhodes2022}), and four orphan afterglows (AT 2019pim, \citealt{Perley2014}; AT 2021any, AT 2021lfa, and AT 2023lcr; \citealt{Li2024}). We show the source size upper limits of these sources from the literature in Figure \ref{sourcesize}, along with that of AT 2023sva, along with its source size upper limit as a function of time, assuming a constant density ISM environment. We see that AT 2023sva has the smallest source size upper limit when compared to the entire population. 
%In addition to these events, GRB 1510127B \citep{Greiner2018} showed large-scale variability in its radio LC, however the variations were too large to be attributed naturally to ISS, and more complex phenomena were invoked to explain the variability. 

%Only AT 2019pim shows a similar source size - $< 6.5 \times 10^{16}$ cm 30 days after explosion, which corresponds to $\sim 13$ days in the rest-frame. This is similar to the source size we derive at a similar time, as we derive a source size limit of $5.2 \times 10^{16}$ cm at 22 days in the rest-frame. 

 The source most similar to AT 2023sva in this parameter space is AT 2019pim, which possesses a source size $< 6.5 \times 10^{16}$ cm 30 days after explosion, which corresponds to $\sim 13$ days in the rest-frame. Figure \ref{sourcesize} suggests that the orphan afterglow population as a whole seems to have smaller source sizes than those of classical GRB afterglows for measurements taken around the same time, as they seem to mark out a different parameter space in the plot. We note that though GRB 210702A and GRB 130427A have smaller source sizes than the rest of the population, this is mainly due to their early-time observations. When extrapolating their source sizes to around the epoch the rest of the sample was observed ($\sim 20$ days), GRB 210702A and GRB 130427A have respective source size upper limits of $10^{18}$ and $7 \times 10^{16}$ cm. 
 
 We then utilize the Anderson-Darling test to determine if the source limits for the classical GRB and orphan afterglow sample are drawn from different statistical distributions. We derive a test-statistic of 4.78, which is greater than the critical value at the 2.5\% significance level of 4.59, with a $p$-value of 0.004. This indicates that the source size upper limits for the sample of classical GRBs and orphan afterglows with ISS source size constraints are drawn from different statistical distributions. We note that the classical GRB population analyzed here is likely a biased population, as ISS analysis is usually only performed for the brightest GRBs.
 
 One possible explanation for this difference is orphan afterglows have lower Lorentz factor origins when compared to classical GRBs, either intrinsically or due to viewing angle effects. Modeling of AT 2020blt and AT 2023lcr showed that they were best modeled as classical GRBs missed by high-energy satellites \citep{Ho2020c, Li2024}. However, \citet{Li2024} showed that the inclusion of early-time data in modeling can lead to different conclusions and neither event had early-time data to constrain the afterglow's rise. AT 2021any was also modeled well as a classical GRB \citep{Gupta2022, Li2024}, though a low-Lorentz factor solution was also proposed \citep{Xu2023}. Modeling of AT 2019pim and AT 2021lfa showed that both low-Lorentz factor solutions and off-axis structured jet solutions were viable \citep{Perley2024, Li2024}, and our analysis of AT 2023sva was not able to place any constraints on the actual nature of the initial outflow at early times, as we determined it could range from highly relativistic to moderately relativistic (see Figure \ref{gamma_in}). 
 
 Therefore, the small observed source sizes for the orphan afterglow population may not be soley due to low-Lorentz factor jets. Some other possibilities are variations in the circumstellar density within the timescale of the observations, small-scale structure contaminating the afterglow images at later times, or stronger scintillation effects than the simple models used for this analysis \citep{Alexander2019}. Further observations of ISS in both orphan  afterglows and classical GRBs are needed in order to understand if the trend holds true for a broader population, and to what extent low-Lorentz factor jets play a role.

\begin{figure*}
    \centering
\includegraphics[width=0.8\linewidth]{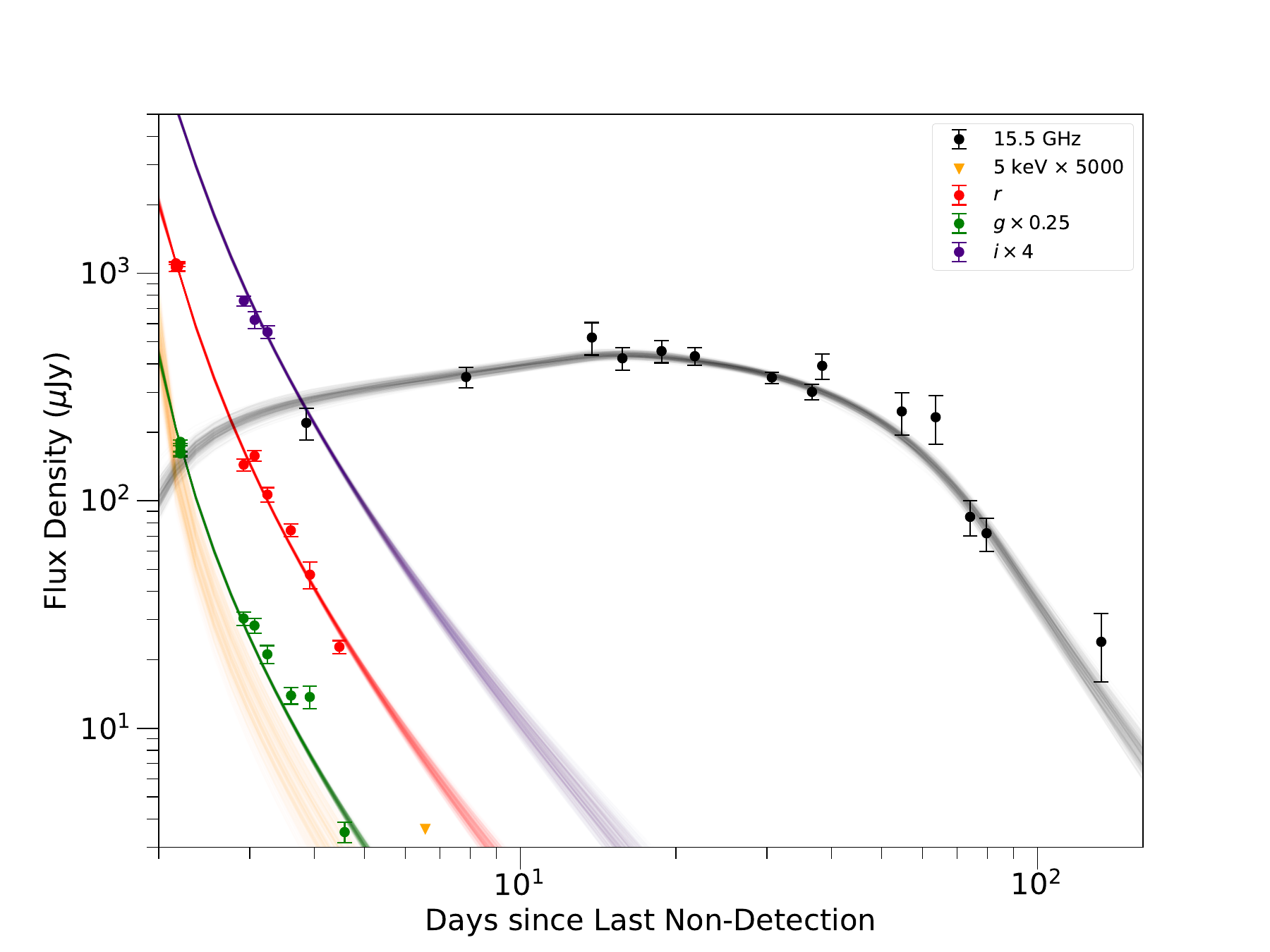}
    \caption{Multiwavelength data set of AT 2023sva, along with the 90\% credible interval of the predicted LCs from our posterior samples for the power-law structured jet model. Fluxes are multiplied by constant factors for visual purposes.}
    \label{bestfit plot}
\end{figure*}

\section{Physical Interpretation of the Outflow}
\label{interpret}

\subsection{Multiwavelength Modeling}
\label{Modeling}

We next model AT 2023sva's LC, utilizing the open-source electromagnetic transient Bayesian fitting software \texttt{redback} \citep{Sarin2024}, to fit different jet models from \texttt{afterglowpy} \citep{Ryan2020}. For the fitting, we utilize the $gri$ optical data, the X-ray upper limit, as well as the 15.5 GHz radio LC. We do not include the lower frequency radio LCs in our fitting procedure as these frequencies are all below $\nu_{\rm{ss}}$ and are affected strongly by ISS. In order to derive posteriors and perform the sampling, we utilize \texttt{bilby} \citep{Ashton2019} and \texttt{Dynesty} \citep{Dynesty}.

We note that \texttt{afterglowpy} has limitations and address the relevant ones individually within the context of AT 2023sva. 
%Support for inverse compton scattering (ICC) is available, though its radiative contribution is overestimated. However, due to the lack of a bright X-ray counterpart to AT 2023sva, the contribution from ICC is negligible for this event, and therefore we do not enable it in our modeling. 
\texttt{afterglowpy} disables jet spreading effects when fixing a finite initial Lorentz factor for the explosion and assumes an infinite initial Lorentz factor when accounting for jet spreading. Because we lack early-time data (two days between the last non-detection and first detection) it is unlikely we will be able to place any constraints on the initial Lorentz factor through modeling. Furthermore, late-time data is largely independent of the initial Lorentz factor, as the jet would have already underwent significant deceleration, so we enable jet spreading effects in our fitting. We note that the corner plots for our modeling presented in the Appendix still show the initial Lorentz factor as a fitting parameter even after enabling jet spreading effects; however, this is just a randomly selected value and has no actual impact on the fitting procedure. 

\texttt{afterglowpy} does not account for synchrotron self-absorption (SSA). SSA is a primarily low-frequency radio phenomenon, and though there is a possibility that the early-time 15.5 GHz LC may be affected by SSA, its high frequency and lack of a prominent LC break at early times implies that SSA effects are negligible at that frequency. 
%Because our radio SEDs only begin at day 73 after explosion, this is far too late in timescale for SSA to be relevant in our observations, and we show that the spectral break $\nu_{\rm{m}}$ has already passed through the radio bands by our observations.  \texttt{afterglowpy} also does not account for any reverse shock (RS) contributions, which arise from a shock propagating back into the shell at around the same time a forward shock is generated from the relativistic outflow interacting with the external medium. The RS primarily impacts emission very shortly after explosion, and the timescale of our observations justifies the assumption of not including a RS component. 
%therefore may impact the early 15 GHz LC, leading to \texttt{afterglowpy} underpredicting the flux. However, the early-time 15 GHz LC is well-fit by a power-law, and there is no clear excess flux that can be attributed to a RS. Therefore, the assumption of not including a RS component in the modeling is justified. 
Finally, \texttt{afterglowpy} does not account for the possibility of a stellar wind medium environment surrounding the blast and assumes a constant density ISM environment. Though it is generally expected that a massive star progenitor should have a stellar wind medium, multiple previous works have shown that a constant density ISM environment still fits well to many LGRBs (though there are exceptions, e.g., \citealt{Panaitescu2001}) -- \citet{Schulze2011} found that out of 27 \textit{Swift} events, two-thirds are compatible with a constant density ISM and \citet{Gompertz2018} found that out of 56 \textit{Fermi} events, half are compatible with a constant density ISM. Furthermore, most GRBs are modeled utilizing a constant density ISM in the literature, making it useful for future comparisons.

We test three different jet structure models from \texttt{afterglowpy} implemented in \texttt{redback}: a tophat jet, a structured jet with a Gaussian profile, and a structured jet with a power-law profile. Tophat jets have a constant energy per unit solid angle, where the bulk of relativistic ejecta is inside the solid angle of the jet. They can be represented by: 
\begin{equation}
E(\theta) = 
    \begin{cases}
        E_{\rm{K,\, iso}}, \, \theta < \theta_{\rm{c}}\\
        0, \,  \theta > \theta_{\rm{c}}
    \end{cases}
\end{equation}
where $E(\theta)$ is the energy with respect to viewing angle, $E_{\rm{K,\, iso}}$ is the isotropic kinetic energy of the jet, $\theta$ is the angle within the jet, and $\theta_c$ is the half-opening angle of the jet's core. The tophat jet has no structure and is the canonical model assumed for most GRB afterglow analyses. However, observations of some GRBs show significant evidence that their jets possess structure (more in \S \ref{comparisonsection}), with weaker emission farther from the jet axis (e.g., \citealt{Troja2019, Cunningham2020,OConnor+2023,Gill2023}). Therefore, we also test the Gaussian structured jet, which is represented by

\begin{equation}
E(\theta) = 
\begin{cases}
       E_{\rm{K,\, iso}}
        (e^{-\theta^2/2\theta_{\rm{c}}^2}), \, \theta  < \theta_{\rm{w}} \\
        0, \,  \theta > \theta_{\rm{w}} 
\end{cases} 
\end{equation}
where $\theta_w$ is a ``wing truncation angle" that represents the relativistic ejecta spreading past the jet's core into wing-like structures. The third model we test is the power-law structured jet, which is represented by

\begin{equation}
E(\theta) = 
\begin{cases}
       E_{\rm{K,\, iso}}\left(1+\Big(\frac{\theta^2}{b\theta_c^2}\Big)^{-b/2}\right),  \theta < \theta_{\rm{w}} \\
        0, \,  \theta > \theta_{\rm{w}} 
\end{cases}
\end{equation}
where $b$ is a power-law index that parameterizes the jet's structure. 

%We note that \texttt{afterglowpy} also has the capability of modeling a guassian jet and power-law jet without an explicit core, instead smoothly varying its structure throughout. However, the parameter estimation from our fitting results were quite similar between these models and the models with their representative core structures. This is because the models without explicit cores are quite flat within the inner regions of the jet, even when the core is not explicitly invoked. Furthermore, most objects in the literature are modeled using the models invoking the core, so we present these models in this work for consistency purposes. 

\begin{figure}
    \centering
    \includegraphics[width=0.99\linewidth]{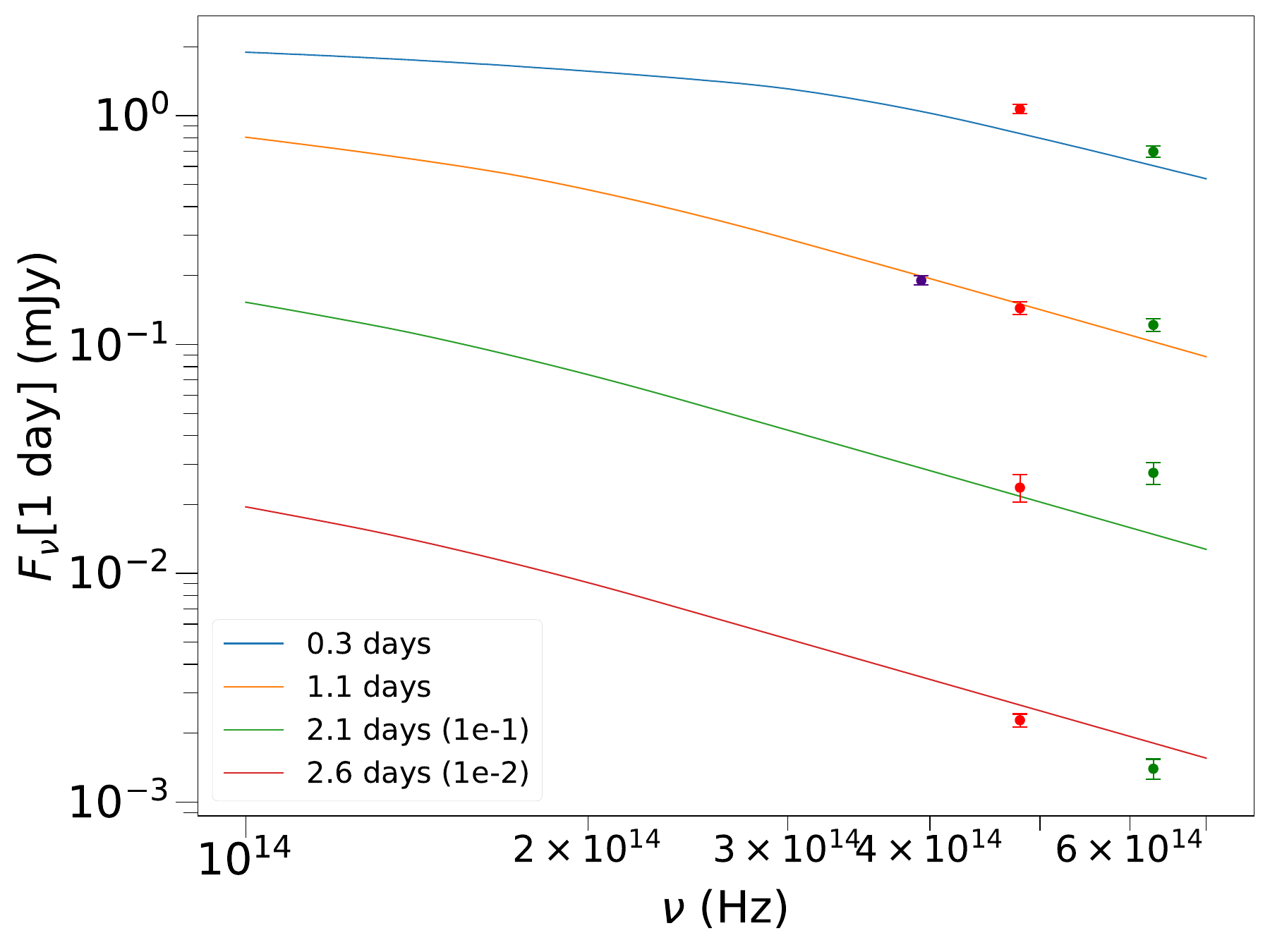}
    \caption{Simulated optical SEDs from the best fit power-law structured jet model and the optical observations. The fluxes are scaled for visual purposes.}
    \label{simulatedspectra}
\end{figure}

\begin{table*}
\caption{Table of priors for \texttt{redback} fits. $\theta_\text{w}$ is ignored by the tophat model and $\beta$ is only used by the power law model.}
\begin{tabular}{cccc}
\hline
Parameter & Unit & Description & Prior (Uniform)\\
\hline
$t_0$ & [MJD] & estimated burst time from first detection& [60202.24, 60204.40] \\

$\theta_\text{v}$ & [rad] & viewing angle & [0, 1.57] \\

$\log_{10}(E_\text{K,iso}/\text{erg})$ & & isotropic equivalent kinetic energy of blast wave along jet axis & [44, 54] \\

$\theta_\text{c}$ & [rad] & half-opening angle of jet core & [0.01, 0.1] \\

$\theta_\text{w}$ & [rad] & wing truncation angle of a structured jet & $[1, 8] \times \theta_\text{c}$ \\

$\log_{10}(n_0/\text{cm}^{-3})$ & & number density of protons in circumburst medium & [$-5$, 2] \\

$p$ & & power law index of relativistic electron energy distribution & [2, 3] \\

$b$ & & power law index of jet angular energy distribution & [0.5, 10]\\

$\log_{10}\epsilon_e$ & & fraction of thermal energy in relativistic electrons & [$-5$, 0]\\

$\log_{10}\epsilon_B$ & & fraction of thermal energy in magnetic field & [$-5$, 0]\\

$\xi_N$ & & fraction of accelerated electrons & [0, 1] \\

\hline
\end{tabular}
\centering
\label{priors}
\end{table*}

We list the priors used for our fitting procedure in Table \ref{priors} and note that we allow for the time of explosion to also be a free parameter, due to the lack of constraints mentioned in \S \ref{ZTFdisc}. We derive the Bayesian evidences for each of the models and find $\rm{log} \, \textit{Z}_{\rm{tophat}} = 15.37$,  $\rm{log} \, \textit{Z}_{\rm{Gaussian}} = 11.92$, and $\rm{log} \, \textit{Z}_{\rm{Power-law}} = 24.41$. We also calculate the Bayesian Information Criterion for every model, which accounts for possible overfitting to the data due to the use of extra parameters in the structured jet models and find $\rm{BIC_{Power-Law}} = -77.63$,  $\rm{BIC_{Gaussian}} = -68.12$, and $\rm{BIC_{Tophat}} = -75.23$, where a lower BIC corresponds to a better fit. Therefore, it is clear the power-law structured jet model is  favored, within the \texttt{afterglowpy} models.

 Furthermore, in both the tophat and Gaussian jet corner plots shown in the Appendix, the posterior for $p$ is hitting the limit $p < 3$, implying that the best-fit $p$ is likely $p > 3$. This is not the case for the power-law structured jet -- though the median $p$ is a high value, it is constrained strongly within the priors of $p = 2.85\pm 0.09$. From our closure relations in \S \ref{closure}, we determined $p = 2.50 \pm 0.14$ in the regime $\nu_{\rm{m}} < \nu < \nu_{\rm{c}}$ -- therefore, the derived $p$ for the tophat and gaussian structured jet models is inconsistent with the $p$ inferred from the modeling. We note that though the relations we used in that section assumed a tophat jet, the relations between $\beta$ and $p$ are identical given a structured jet \citep{Ryan2020}. This gives further evidence that a power-law structured jet is favored.

We present the 90\% credible interval of the predicted LCs from the posterior samples of the power-law structured jet model, along with the full multi-wavelength set of observations, in Figure \ref{bestfit plot}. We note that this is the preferred model in the context of the \texttt{afterglowpy} model and without considerations for any more complex physics contributing to the LC and spectra. Therefore, we stress that though AT 2023sva shows significant evidence for possessing a power-law structured jet within this context, we cannot definitely claim that this is the only possible scenario. 

%From our modeling, we find that a tophat jet and a Gaussian structured jet cannnot reproduce our multiwavelength observations to the same degree that the power-law structured jet can. Though the tophat and Gaussian jets are able to reproduce the optical LCs as well as the power-law model (see left hand side of Figure \ref{modeling}), they cannot reproduce the behavior of the radio LCs (see right hand side of Figure \ref{modeling}) as well, as they are not able to capture the initial rise to shallow decay to steep decay that is observed. However, the differences between the models are subtle, and hard to tell by eye. 

%From \ref{radioLC}, we see that the 15.5 GHz LC first transitions to a shallow decay after the peak flux, evolving with $t^{-0.4}$, and then eventually transitions to a steeper decay at later times, evolving with $t^{-2}$. Both the top-hat and gaussian stuctured jet models Only the power-law model is able to reproduce this behavior. We do note that the power-law model does overpredict the flux slightly at very early-times, with the first observation.  The peak flux for both of the models occurs at a later time that observed in the radio LC (see Figure \ref{radioLC}), and the LC transitions rapidly after the peak flux to a steep power-law decay. 

\begin{table}
\caption{Final best-fit and median $\pm \, 1\sigma$ parameters for the power-law structured jet model, which is the most favored model for AT 2023sva. The best-fit parameters correspond to the model that possesses the maximum log likelihood and the corner plot corresponding to median parameters and their 1$\sigma$ confidence intervals are presented in the Appendix.}
\centering
\begin{tabular}{lll}
\hline
Parameter & Best-fit Result & Median $\pm \, 1\sigma$\\
\hline
$t_0$ [MJD] & $60204.09$ & $60204.09^{+0.03}_{-0.03}$\\
$\theta_\text{v}$ [rad] & $0.08$ & $0.07 \pm 0.02$\\
$\log_{10}(E_\text{K,iso}/\text{erg})$ & $53.63$ & $53.33^{+0.27}_{-0.21}$\\
$\theta_\text{c}$ [rad] & $0.05$ & $0.06 \pm 0.02$\\
$\theta_\text{w}$ [rad] & $7.71 \times \theta_{\rm{c}}$ & $6.81^{+0.79}_{-1.01} \times \theta_{\rm{c}}$\\
$\log_{10}(n_0/\text{cm}^{-3})$ & $1.41$ & $1.40^{+0.40}_{-0.74}$\\
$p$ & $2.84$ & $2.85\pm 0.09$\\
$b$ & $1.02$ & $0.99^{+0.36}_{-0.23}$\\
$\log_{10}\epsilon_e$ & $-0.77$ & $-0.53^{+0.18}_{-0.28}$\\
$\log_{10}\epsilon_B$ & $-3.32$ & $-3.18^{+0.45}_{-0.26}$\\
$\xi_N$ & $0.42$ & $0.70^{+0.19}_{-0.27}$\\
\hline
\end{tabular}
\label{bestfit}
\end{table}

We then compute the angular size of the best-fit power-law afterglow model's image on the sky, as a function of frequency and time, to compare it to the size derived from the ISS of the observed LC described in \S \ref{scintillation}. We find that at 73 days at 15.5 GHz, the source has an angular size of $\theta \approx 1.95 \, \rm{\mu arcseconds}$. This is consistent with the  ISS analysis (\S \ref{scintillation}), where we determined that AT 2023sva's angular size needed to be comparable or smaller than the Fresnel scale (2 $\mu$arcsec). This gives us an independent confirmation that our modeling results match well with the radio observations. From the best-fit parameters shown in Table \ref{bestfit}, we see that we are viewing the power-law structured jet slightly off-axis, where $\theta_{\rm{v}} \gtrsim \theta_{\rm{c}}$. Furthermore, the power-law index of the energy distribution as a function of viewing angle is quite shallow, as we derive $b = 1.02$ for the best-fit model and a median $\pm \, 1\sigma$ is $0.99^{+0.36}_{-0.24}$. We discuss the implications of this in \S \ref{comparisonsection}.

In Figure \ref{simulatedspectra}, we show the simulated optical SEDs generated by \texttt{afterglowpy} utilizing the best-fit power-law structured jet model, along with the optical observations of AT 2023sva. We see that generally the simulated SEDs match well with the observations across all epochs. Furthermore, we see there is a clear break in the optical SED in the 0.3 day epoch at around $4 \times 10^{14}$ Hz and this break moves further down in frequency over time. This is the cooling break $\nu_{\rm{c}}$. This implies that the optical bands lie just above the cooling break, which is at odds with our analysis in \S \ref{closure}, where we determined that the optical bands should lie below the cooling break to imply a physical $p$. This discrepancy between the two analyses can be explained by the fact that spectral breaks are smooth over orders of magnitude. Because the optical bands have barely passed $\nu_{\rm{c}}$ at the time the spectral index was derived ($\sim$ 2 days), the slope has not fully steepened yet, leading to a fulfillment of the relation in the $\nu_{\rm{m}} < \nu < \nu_{\rm{c}}$ regime. Furthermore, Figure \ref{opticalLC} shows evidence of a possible temporal break in the early-time LC, which would correspond to $\nu_{\rm{c}}$ fully passing through the optical bands.

Given that the best-fit model and posteriors indicate that we are viewing the jet slightly off-axis just outside the jet's core,  ($\theta_{\rm{v}} \gtrsim \theta_{\rm{c}}$), we utilize the closure relations provided by \citet{Ryan2020} for a structured jet for misaligned viewers, to check if the relationship between $\alpha$, $p$, and $b$ hold true. We test the relations in two different regimes: for $\nu_{\rm{m}} < \nu < \nu_{\rm{c}}$ the relation is  $\alpha =
(3-6p+3g)/(8+g)$ and for $\nu_{\rm{m}} < \nu_{\rm{c}} < \nu$, the relation is $\alpha = (1-6p+2g)/(8+g)$. The $g$ parameter accounts for the angular structure of the jet in the relations and is given by

\newcommand{\thobs}{{\theta_{\mathrm{obs}}}}
\newcommand{\thC}{{\theta_{\mathrm{c}}}}
\newcommand{\theff}{{\theta_{\mathrm{eff}}}}

\begin{align}
	g &= \frac{2 b (\thobs-\theff)\theff}{b \thC^2+\theff^2} \label{eq:geff},
\end{align}
where for a power-law structured jet, $\theta_{\rm{eff}}$ is 

\begin{align}
	\theff &= \thobs \big [1.8+2.1b^{-1.25}  \nonumber  \\
	 	& \qquad + (0.49-0.86b^{-1.15}) \thobs/\thC \big ] ^{-1/2} &&\label{eq:theff}.
\end{align}

Utilizing the best-fit values from the modeling, we find $\theff = 0.044$ and $g = 0.72$. Applying the closure relations, we find $\alpha \sim 1.4$ for $ \nu_{\rm{m}} < \nu < \nu_{\rm{c}}$ and $\alpha \sim 1.7$ for $\nu_{\rm{m}} < \nu_{\rm{c}} < \nu$. In \S \ref{opticalLCsection}, we found $\alpha = 1.64 \pm 0.02$, which matches well with the $\nu_{\rm{m}} < \nu_{\rm{c}} < \nu$ case -- and from the simulated spectra in Figure \ref{simulatedspectra}, we know that the optical bands lie in this regime. This gives further evidence favoring the power-law structured jet model.

%\citet{Racusin2009} provides a closure relation between the temporal power-law decay index ($\alpha$), spectral index ($\beta$), and power-law index for a structured jet ($b$), in the two regimes of $\nu_{\rm{m}} < \nu < \nu_{\rm{c}} $ ($\alpha = \frac{12\beta + 3b}{8 - b}$), and $\nu_{\rm{m}} < \nu_{\rm{c}} < \nu $ ($\alpha = \frac{12\beta + 2b - 4}{8 - b}$) in a constant density ISM.  Utilizing $b = 1.02$ from our multiwavelength modeling, and $p  = 2.84$ from \S \ref{SEDoptical} (which implies $\beta = 0.92$ for $\nu_{\rm{m}} < \nu < \nu_{\rm{c}} $ and $\beta = 1.42$ for $\nu_{\rm{m}} < \nu_{\rm{c}} < \nu $), we find $\alpha \approx 2.0$ for $\nu_{\rm{m}} < \nu < \nu_{\rm{c}} $ and $\alpha \approx 1.3$ for $\nu > \nu_{\rm{c}}$. The $\nu_{\rm{m}} < \nu < \nu_{\rm{c}} $ case is consistent with the power-law decay index derived in \S \ref{opticalLCsection}. Therefore, the closure relations between $\alpha$ and $\beta$ also follow a similar pattern to the relations between $\beta$ and $p$, where the relations satisfy the regime $\nu_{\rm{m}} < \nu < \nu_{\rm{c}} $ even though physically,  $\nu_{\rm{m}} < \nu_{\rm{c}} < \nu $. Again, this discrepancy is due to the proximity of the optical bands to the cooling break. We note that the $\beta$ implied from the modeling is steeper than the $\beta$ derived from fitting the optical spectrum ($\beta = 0.75 \pm 0.07$) for the fulfilled relation - but still consistent within 3$\sigma$. 

\subsection{Why the lack of $\gamma$-rays?}
The question arises as to why we did not detect any associated $\gamma$-ray emission for AT 2023sva, at least to an upper limit of $E_{\gamma, \rm{iso}} < 1.6 \times 10^{52} \, \rm{erg}$. Studies suggest that for structured jets, prompt $\gamma$-ray emission may only be efficiently produced in the core of the jet \citep[e.g.,][]{OConnor+2023, Beniamini2019, Beniamini2020, Gill2020}. Since the Lorentz factor of a structured jet will also decrease with angle, this leads to an increase in the opacity due to photon-pair production processes \citep{Gill2020}. A lower Lorentz factor along the line of sight will also decrease the dissipation radius of the prompt emission, so that at large viewing angles, the photospheric radius is larger than the dissipation radius, resulting in high optical depth regions \citep{LambKobayashi2016,Beniamini2020}. These effects are highly dependent on the angular Lorentz factor profile and the initial core Lorentz factor. 

Both of these effects suppress the $\gamma$-ray emission \citep{Gill2020} and decrease the observed $\gamma$-ray efficiency, which describes how efficiently the jet converts its energy to radiation along the line of sight. Therefore, structured jets need to be viewed extremely close to on-axis to detect their associated $\gamma$-ray prompt emission \citep[e.g.,][]{OConnor2024}, even if the viewing angle is within the wing truncation angle. Through our  modeling analysis, the posteriors indicate $\theta_v = 0.07 \pm 0.02$ and $\theta_c = 0.06 \pm 0.02$. Therefore, we are likely viewing AT 2023sva slightly off-axis, making this scenario a strong possibility to explain the lack of $\gamma$-rays, depending on the steepness of the angular Lorentz factor profile and initial core Lorentz factor. 
 %If the event is being viewed slightly off-axis at or just outside the jet's core opening angle, with $\theta_v \gtrsim \theta_c$, this supports the conclusion that AT 2023sva possesses a structured jet, as viewing structured jets on the edge or slightly outside their core could result in missing $\gamma$-ray emission \citep{ Nakar2003,Rossi2008, Cenko2013,Salafia2015, Gavin2017,Gavin2018, Huang2020, Sarin2021, OConnor+2023, Freeburn2024}. 

We then quantify the $\gamma$-ray radiative efficiency of AT 2023sva. The radiative efficency is calculated through 
\begin{equation}
    \eta_\gamma = \frac{E_{\gamma,\text{iso}}}{E_\text{K,iso} + E_{\gamma,\text{iso}}}.
\label{eq:eff}
\end{equation}
%If close to on-axis, low radiative efficiency can cause a lack of observed $\gamma$-rays \citep{Sarin2022}. 
Using the upper limit for $E_{\gamma, \rm{iso}}$ and the 1$\sigma$ range of $E_{\rm{K,iso}}$ from the best-fit power-law structured jet model, we derive $\eta_\gamma < 4 - 11\%$. 
These efficiencies are on the lower end of the distribution observed from LGRBs \citep{Wang2015}, but are consistent with the $\sim 1\%$ efficiency derived from the internal shock model used to describe the prompt emission of bursts \citep{Kumar1999}. 
Several other orphan afterglows in the literature have demonstrated possible low efficiencies -- AT 2020blt ($< 0.3 - 4.5\%$;  \citealt{Sarin2022}, $< 0.2 - 17.9\%$; \citealt{Li2024}), AT 2020lfa ($< 0.01 - 0.05\%$; \citealt{Ye2024}) and AT 2023lcr ($< 1.3 - 3.4\%$; \citealt{Li2024}), which provides evidence that the lack of associated $\gamma$-ray emission for at least some orphan afterglows may be attributed to their low radiative efficiencies. 

It is important to note that these efficiencies are calculated along the line of sight, so if viewed off-axis, low efficiencies are a natural consequence of structured jets. It is also possible that on-axis jets may have intrinsic low efficiencies, as suggested for AT 2020blt in \citet{Sarin2022}. It is more likely that AT 2023sva's low efficiency is due to viewing the event slightly off-axis; however, an intrinsic low efficiency cannot be ruled out, as the posteriors from the modeling do not indicate an extremely off-axis viewing angle.

%calculatedHowever it is not clear whether AT 2023sva is in that regime, as efficiencies between 10 and 14\% are commonly seen in LGRBs \citep{Wang2015}, and it is likely we are viewing this event slightly off-axis. 

%Another possible explanation for the lack of associated $\gamma$-rays is AT 2023sva's shallow structured jet. \citet{OConnor+2023} suggested that for shallow structured jets, prompt $\gamma$-ray emission may only be produced in the core of the jet, as the Lorentz factor of a structured jet will decrease with angle. This leads to an increase in the opacity to photon-pair production processes, suppressing $\gamma$-ray emission Therefore, there is a need to be \textit{extremely} on-axis to these shallow structured jets to detect their associated $\gamma$-ray prompt emission, as photon pair-production processes will stifle $\gamma$-ray emission at wider viewing angles similar to a dirty fireball, even if the viewing angle is within the wing truncation angle. \citet{OConnor+2023} state that this process would lower the the total energy released in $\gamma$-rays, or $E_{\rm{\gamma, \, iso}}$, by a factor of more than 20, though a radiative efficiency $\eta_\gamma \gtrsim 20\%$ is necessary. However, for AT 2023sva, we calculate  $\eta_\gamma < 28\%$, and $\theta_c \sim \theta_v$. Therefore, we determine that the most likely reason that AT 2023sva was not detected in $\gamma$-rays is due to its low $\gamma$-ray efficiency. 
\begin{figure}
    \centering
    \includegraphics[width=0.99\linewidth]{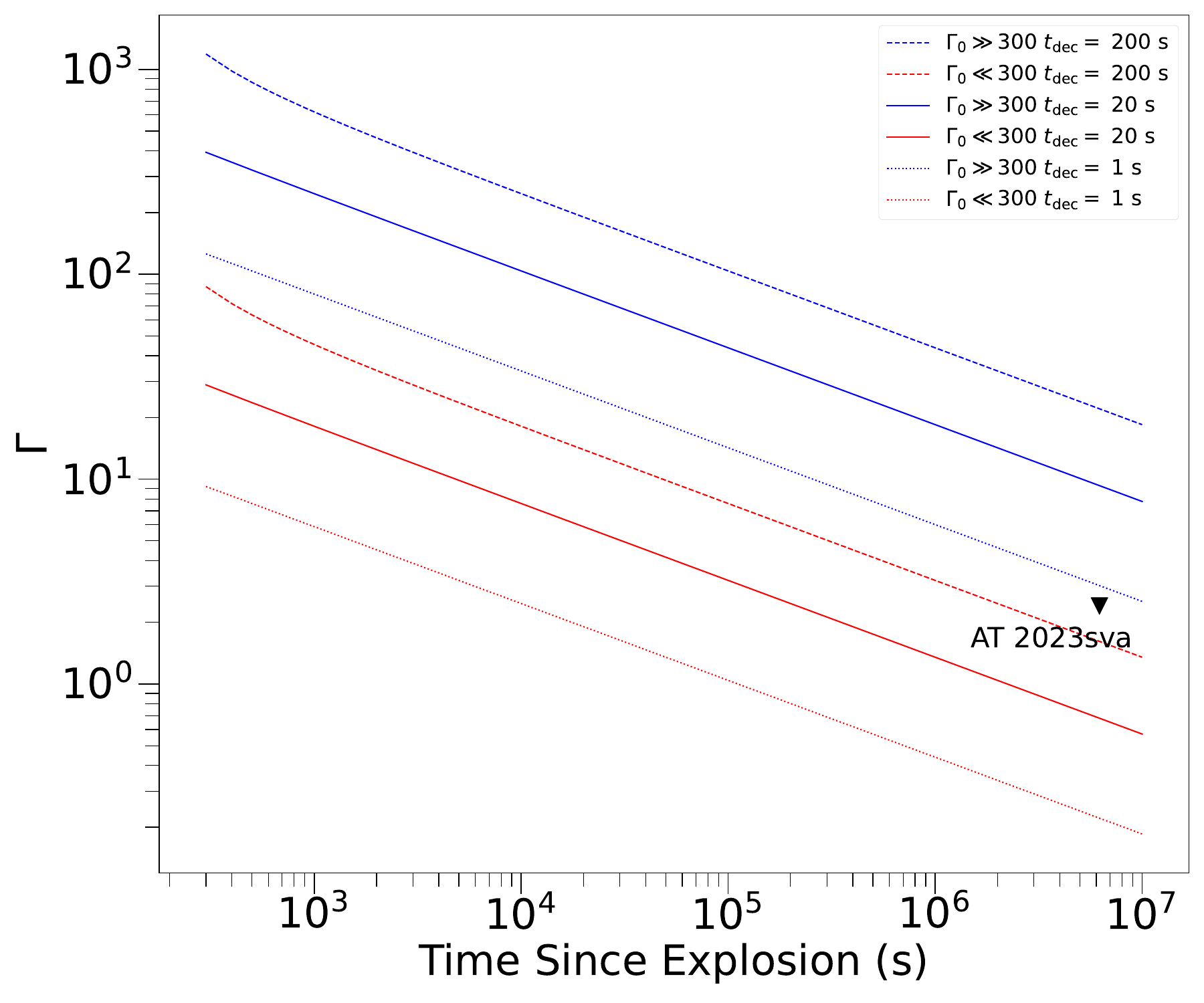}
    \caption{Comparison of the Lorentz factor evolution of $\Gamma \gg 300$ and $\Gamma \ll 300$ jet models (from \citealt{Dermer1999}) with respect to different observed deceleration times. We also show the upper limit for AT 2023sva's Lorentz factor at the observed time of 72 days from the scintillation analysis ($\Gamma_{\rm{av, 72 \, d}} < 2.4$). }
    \label{fireballmodel}
\end{figure}

If the outflow is intrinsically moderately relativistic in its core, a low Lorentz factor in comparison to classical LGRBs ($\Gamma > 100$) could be a possible reason for the lack of associated $\gamma$-rays, as pair production processes can reprocess $\gamma$-rays to X-rays if the jet is baryon-loaded, leading to a low Lorentz factor. It has been shown that even jets with moderately high Lorentz factors ($\Gamma \sim 50$) can have their high-energy emission suppressed \citep{LambKobayashi2016, Matsumoto2019}. This cannot be ruled out for AT 2023sva, as its small source size at late times (see \S \ref{scintillation}) gives possible evidence for this case.

Therefore, we try to measure the initial Lorentz factor $\Gamma_0$ through the relationship between $\Gamma_0$, $t_{\rm{dec}}$, $E_{\rm{K, \, iso}}$, and the circumburst medium density $n_0$, in the case of a constant density ISM environment. This is given by 
\begin{equation}
\Gamma_{2.5} = \left( \frac{10\,\text{s} (1 + z)}{t_\text{dec}} \right)^{3/8} \left(\frac{E_{53}}{n_0}\right)^{1/8},
\label{eq:tdec}
\end{equation}
from \citet{Meszaros_2006}, where $n_0$ is the number density of the circumburst medium in cm$^{-3}$, $E_{53} = E_{\text{K,iso}} / 10^{53}$ erg, and $\Gamma_{2.5} = \Gamma_0 / 10^{2.5}$. We first derive a lower limit, assuming that $t_{\rm dec}$ is at the time of first detection, or 7.5 hours after the explosion in the observer frame (see Figure \ref{gamma_in} for how different assumptions of $t_{\rm dec}$ impact $\Gamma_0$). Using the full, weighted posterior distributions of $E_{\rm{K, \, iso}}$ and $n_0$ from modeling the power-law structured jet, we derive a lower limit on the initial Lorentz factor of $\Gamma_0 > 16^{+5}_{-3}$. If we assume that $t_{\rm dec}$ is 20 seconds in the rest frame as we did in our radio ISS analysis (see \S \ref{scintillation}), we find $\Gamma_0 = 182^{+49}_{-27}$. This is consistent with the upper limit derived in \S \ref{scintillation} assuming the same $t_{\rm dec}$, where $\Gamma_0 < 178$. Again, we note that the assumed $t_{\rm{dec}}$ is a major caveat, as past GRBs have derived values between 1 and 1000 seconds \citep{Ghirlanda2018}. If $t_{\rm dec} \gtrsim 94$ seconds, that would imply $\Gamma_0 \lesssim 102^{+27}_{-15} $, meaning that AT 2023sva possesses ejecta more moderately relativistic than the classical GRB population. This is consistent to the limit derived in \S \ref{scintillation}, where we found if $t_{\rm dec} \gtrsim 94$ seconds, that would imply $\Gamma_0 \lesssim 100 $ as well.

We then test the physical models of \citet{Dermer1999} in Figure \ref{fireballmodel}. \citet{Dermer1999} introduce functions that model the synchrotron emission produced by relativistic blast waves driven from GRB jets with $\Gamma_0 \gg 300$ and $\Gamma_0 \ll 300$, that track the Lorentz factor evolution with time. \citet{Dermer1999} make the distinction between clean and dirty fireballs for these two different jets; however, true dirty fireballs are now characterized by $\Gamma \lesssim 10$. Therefore, we perform this analysis not to characterize AT 2023sva as a true dirty fireball or not, but to provide another independent constraint of the Lorentz factor. In order to do this, we utilize Equations 4, 6, 11, 27, and 30 from \citet{Dermer1999}, which are sensitive to $n_0$, $\epsilon_e$, and $\epsilon_b$, which we constrain to our best-fit values from \S \ref{Modeling}. These equations are also sensitive to the observed  $t_{\rm{dec}}$ and we show a range of values for $t_{\rm{dec}}$ in Figure \ref{fireballmodel}, along with the Lorentz factor upper limit derived from the source size in \S \ref{scintillation}. 

This analysis indicates that AT 2023sva originates from a jet with $\Gamma \ll 300$, which is consistent with our previous analyses. This modeling is very sensitive to the choice of $n_0$, $\epsilon_e$, and $\epsilon_b$, which is affected by the caveats associated with our afterglow modeling in \S \ref{Modeling}, so it is difficult to make any more robust conclusions. Overall, through all the independent methods used to constrain $\Gamma_0$, we cannot rule out that AT 2023sva has a classical, high $\Gamma$ origin, and we find that the lack of associated $\gamma$-rays is most likely due to viewing a structured jet slightly off-axis -- though we also cannot rule out an on-axis, low radiative efficiency burst, or a more moderately relativistic outflow than classical GRBs. Further constraints on $t_{\rm{dec}}$ would allow us to place more robust constraints on $\Gamma_0$, highlighting the importance of high-cadence early-time observations in orphan afterglow observations.

\begin{figure}
    \centering
    \includegraphics[width=0.99\linewidth]{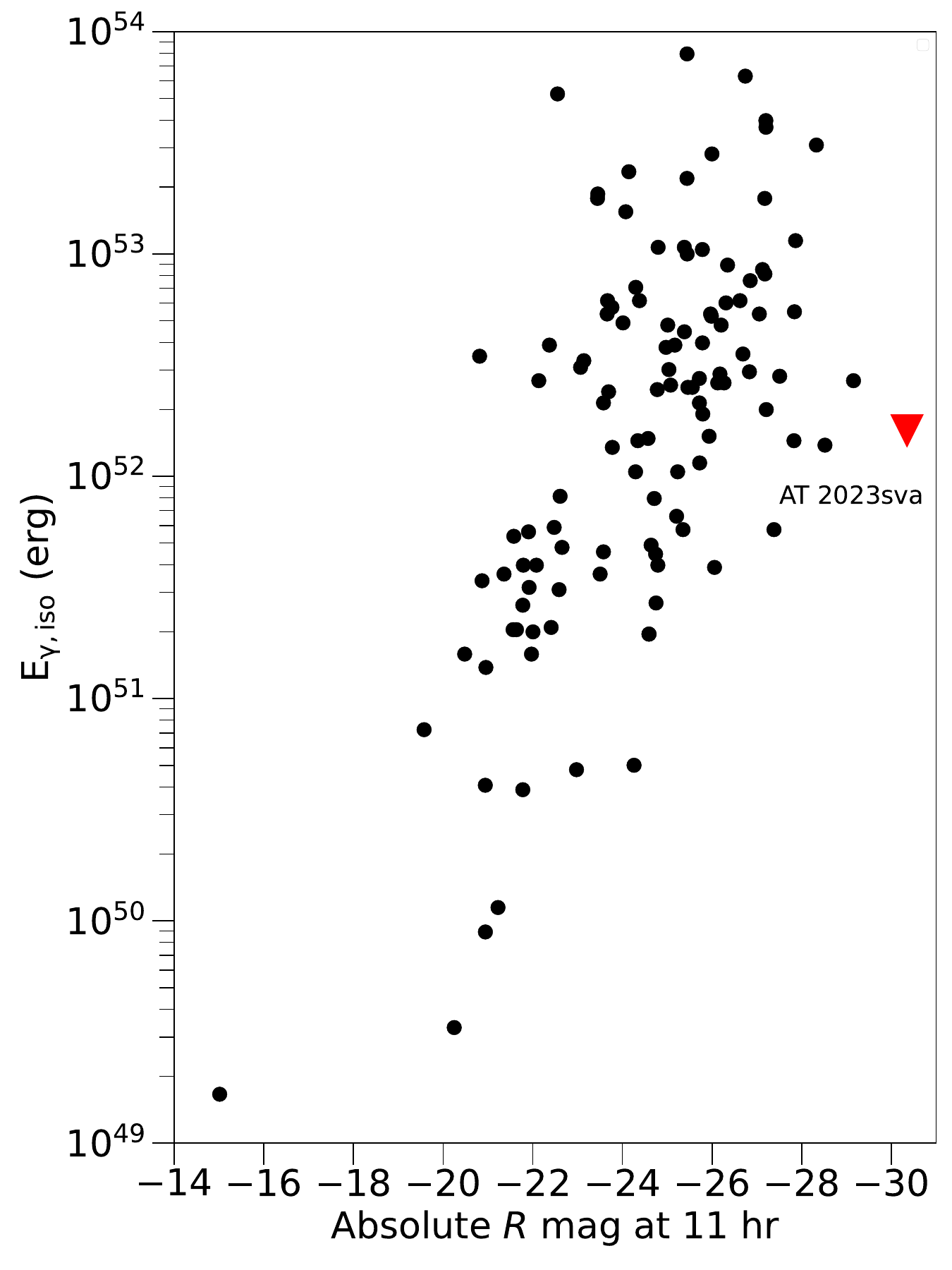}
    \caption{$E_{\rm{\gamma, \, iso}} $ plotted against the absolute $r$-band optical magnitude at 11 hours for GRBs in the literature (from \citealt{Nysewander2009}) along with AT 2023sva, shown in red.}
    \label{grbcomparison}
\end{figure}

%This analysis independently agrees with the analysis from \S \ref{scintillation}, in that the outflow can range from being highly relativistic to moderately relativistic. 

%However, the lack of more constraining $\gamma$-ray limits, late-time detections to constrain the post jet-break evolution, and a detected X-ray counterpart, limit our ability to truly understand the properties of AT 2023sva's initial outflow.

%Does this make AT 2023sva a possible dirty fireball? It could be the case -- however, our observations are not constraining enough to determine whether it is or is not with certainty. 
%It has been shown that GRBs with initial Lorentz factors as low as $\Gamma_0 = 20$ can produce weak $\gamma$-ray prompt emission \citep{Ghirlanda2018}, so it is possible that AT 2023sva produced $\gamma$-rays that were missed by high energy satellites, given our constraints on $E_{\rm{\gamma, \, iso}}$. 

\subsection{Comparison to GRB Afterglows}
\label{comparisonsection}
In Figure \ref{grbcomparison}, we show AT 2023sva in the context of other LGRBs in the literature \citep{Nysewander2009} that have measurements of $E_{\rm{\gamma, \, iso}}$ and an optical afterglow detection. We use the upper limit of $E_{\rm{\gamma, \, iso}} < 1.6 \times 10^{52}$ erg derived in \S \ref{grbsearch}. The optical magnitudes are all normalized to 11 hours after explosion in every GRB's respective rest frame, in the rest-frame $r$ band. Extrapolating our first observed $r$ band measurement (corrected for galactic and host-galaxy extinction) to 11 hours after explosion and utilizing $\beta = 0.75$ (see \S \ref{SEDoptical}) to transform to the rest-frame $r$ band, we find $M_{{\rm{abs, \,}}r}  \sim -30.4$ mag, making it the most luminous optical afterglow in the sample. 

%However, AT 2023sva's actual location in the two-dimensional parameter space is not particularly striking, as the events around its absolute magnitude show quite some scatter with respect to their $E_{\rm{\gamma, \, iso}}$. There are a handful of events with a higher  $E_{\rm{\gamma, \, iso}}$ at similar optical luminosities, and also lower $E_{\rm{\gamma, \, iso}}$  at similar optical luminosities.

\begin{figure}
    \centering
    \includegraphics[width=0.99\linewidth]{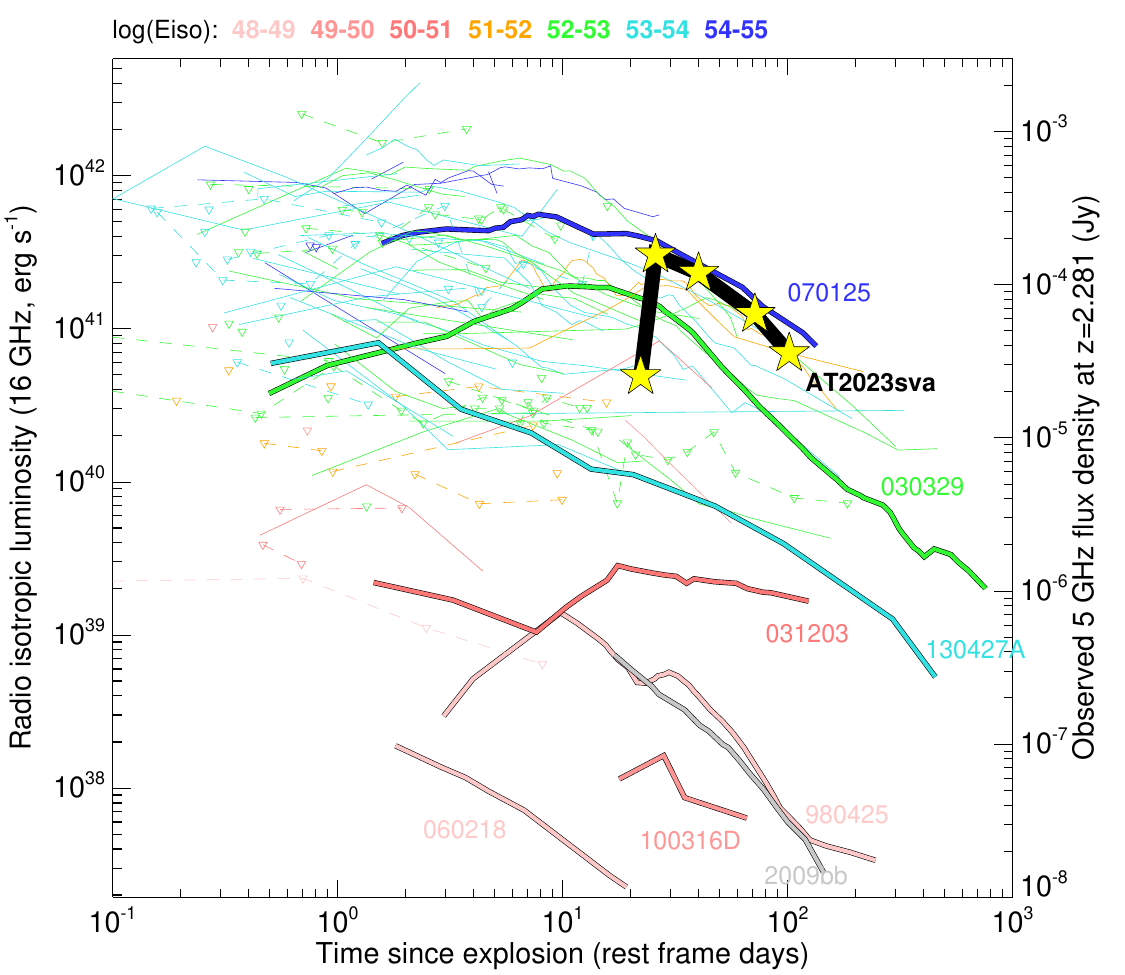}
    \caption{Comparison of AT 2023sva's radio LC (stars) to other GRB afterglow LCs from literature \citep{Chandra2012, Perley2014}. We show the rest-frame 16 GHz luminosity LC, which corresponds to an observed flux density LC at 5 GHz at a redshift $z = 2.28$. Events are also color-coded with respect to their $E_{\rm{\gamma,\, iso}}$, except fot AT 2023sva.}
    \label{radiocomparison}
\end{figure}

In Figure \ref{radiocomparison}, we show AT 2023sva's radio LC compared to other GRB afterglows in the literature (plot modified from \citealt{Chandra2012, Perley2014}). The LCs are plotted with respect to their 16 GHz rest-frame luminosity, which corresponds to a flux density at 5 GHz at $z = 2.28$. The flux densities were converted to luminosities through multiplying by the distance luminosity associated with $z = 2.28$. The plot is also color-coded with respect to the GRBs' associated $E_{\rm{\gamma,\, iso}}$. The large rise between the first two epochs is due to ISS, as the redshift of AT 2023sva places it in a regime where the rest-frame 16 GHz LC is affected heavily by ISS. We see that AT 2023sva is quite radio-loud and most GRBs that have a similar radio luminosity have $E_{\rm{\gamma,\, iso}} > 10^{52}$ erg. However, in \S \ref{grbsearch} we ruled out a GRB counterpart to an upper limit of $E_{\rm{\gamma,\, iso}} < 1.6 \times 10^{52}$ erg. This makes AT 2023sva a unique event in this parameter space. 

Now, we visit AT 2023sva's preference for a structured jet model within the context of GRB afterglow jets from the literature. Structured jets have been proposed as possible models for GRB afterglows for some time, as it is natural that a jet with structure should develop due to breaking out of a dense stellar environment \citep[e.g.,][]{Gottlieb2021, Gottlieb2022}. However, there have only been a few possible structured jet models inferred through afterglow analyses in the literature. GRB 030329 showed two different jet breaks -- one in the optical at around 0.55 days and another in the radio at around 9.8 days, leading to an interpretation of its afterglow as a two-component structured jet \citep{Berger2003}, though alternative models were not excluded. GRB 130427A's X-ray LC was suggested to arise from a power-law structured jet \citep{DePasquale2016}, though other models were also discussed. GRB 160625B showed strong evidence for possessing a Gaussian-structured jet, when compared to a tophat jet or power-law structured jet \citep{Cunningham2020}. GRB 190828A was also suggested to have a two-component jet model, to discriminate between the early X-ray and optical emission and late-time X-ray and radio emission \citep{Sato2021}. 

Finally, the two most famous examples of structured jets are in GRB 170817A and GRB 221009A. GRB 170817A was a short GRB that also had associated gravitational waves (e.g., \citealt{GRB170817A}), making it one of the most extensively followed-up astrophysical sources in history. Through extensive modeling efforts across the electromagnetic spectrum, it became clear that a structured jet was the only model that could explain the multiwavelength observations (both power-law and Gaussian structured jets are presented as viable models in various works; e.g., \citealt{Margutti2018, Troja2018, Lazzati2018, Ryan2024}). GRB 221009A was the brightest GRB ever detected with respect to its fluence \citep[e.g.,][]{Lesage+2023,Frederiks+2023} and it is one of the most energetic GRBs ever detected \citep{Burns2023}. \citet{OConnor+2023}, \citet{Gill2023}, \citet{LHAASO+2023}, and \citet{Rhodes2024} suggest that it possesses a shallow structured jet best modeled using a broken-power law energy distribution. As an alternative explanation a two-component jet model was also proposed by \citet{Sato2023}. 

In this work, we show evidence that AT 2023sva should be added to the above list and it is the third orphan event to show possible evidence for possessing complex jet structure. The other two are AT 2019pim \citep{Perley2024} and AT 2021lfa \citep{Li2024}, as both of these events are modeled well by both low-Lorentz factor solutions or off-axis structured jet solutions. AT 2023sva's shallow power-law index for its structured jet's energy distribution with respect to viewing angle ($b = 0.99^{+0.36}_{-0.24}$) is similar to the index derived by \citet{OConnor+2023} for GRB 221009A, as they derived indices of 0.75 and 1.15 for their broken-power law distribution, and \citet{Gill2023} who found an index of 0.8. However, the energetics of GRB 221009A and AT 2023sva are quite different -- GRB 221009A possessed an $E_{\rm{\gamma, iso}} \approx 10^{55}$ erg \citep{Lesage+2023} and had strong X-ray emission. This gives evidence that the simplified jet model assumed for GRBs may need to be revised, as one of the most energetic, explosive GRBs ever detected along with an orphan event both show evidence for possessing shallow structured jets.

\section{Summary and Conclusion}
\label{conclusion}
In this work, we present the discovery of an orphan  afterglow, AT 2023sva, at a redshift $z = 2.28$. We analyze the optical, X-ray, and radio observations of AT 2023sva and place it in the context of GRBs in the literature. Our main findings are:

\begin{itemize}
    \item AT 2023sva does not possess an associated GRB counterpart, based on a search of $\gamma$-ray satellite archives between the last non-detection and first detection of AT 2023sva (a two day window), to an isotropic equivalent energy upper limit of $E_{\gamma, \, \rm{iso}} < 1.6 \times 10^{52} $ erg. 
    \item There is additional host-galaxy extinction present ($E_{\rm{B-V, \, host}}  = 0.09 \pm 0.01$), due to a characteristic curvature in the optical spectrum. Only a few absorption features are present in the spectrum and their line strengths are weaker than 99\% of GRBs in the literature.

    \item The radio LC and SED shows clear presence of interstellar scintillation 72 days after the explosion in the observer frame. We use this to provide an upper limit constraint on the bulk Lorentz factor in the rest frame of $\Gamma_{\rm{av, 22.0 \, days}} \leq 2.4$ and extrapolating back to the time of first detection, we derive a constraint of $\Gamma_{\rm{av, 2.3 \, hr}} \leq 19$. A more stringent constraint on the initial Lorentz factor was not able to be determined, due to a lack of earlier-time observations.
    
    \item AT 2023sva has a small source size upper limit ($5.2 \times 10^{16}$ cm) derived from ISS constraints when compared to most classical GRBs and shares scintillation properties in common with other orphan afterglows. We determine that the orphan afterglow population has statistically lower source-size upper limits than the classical GRB population, for events whose limits were derived from ISS analyses.
    
    \item The model that can best reproduce the multi-wavelength observations is a slightly off-axis ($\theta_{\rm{v}} \gtrsim \theta_{\rm{c}}$) shallow power-law structured jet, which we determined through Bayesian multi-wavelength modeling of the afterglow. We only model the source in a constant density ISM and cannot rule out the possibility of the source originating from a wind medium. AT 2023sva's shallow jet structure is remarkably similar to that of GRB 221009A \citep{OConnor+2023, Gill2023, LHAASO+2023, Rhodes2024}.
    
    \item The lack of a detected associated GRB counterpart is most likely due to viewing the structured jet slightly off-axis, just outside the opening angle of the jet's core. However, this is not the only possibility and we determine that it may be due to its lower radiative efficiency, or possibly a more moderately relativistic outflow than classical GRBs.
\end{itemize}

 Because the early evolution of afterglow LCs depends strongly on the jet's structure, it is vital that orphan afterglow searches in the future calibrate their observation strategies with more complex structured jet models (e.g., \citealt{Lamb2017, Lamb2018, Xie2019,  Freeburn2024}). This is incredibly important in the coming years, as new instruments like the Vera Rubin Observatory \citep{LSST} will vastly increase the discovery space of orphan afterglows, due to increased sensitivity. However, this discovery space can only be utilized if observing strategies are broadened to incorporate the diverse range of angular energy profiles of GRB jets. 

Furthermore, the recently launched Einstein Probe (EP; \citealt{Yuan2022}) and the Space-based multi-band astronomical Variable Objects Monitor (SVOM; \citealt{Wei2016}) will increase the number of afterglows detected without associated $\gamma$-ray emission, through discovering their X-ray ``prompt" emission, providing another avenue for characterizing these events. In fact, EP has already began detecting GRB-related events in the soft X-rays. One such event, EP 240414a, did not have significant associated $\gamma$-ray emission \citep{Bright2024} and was followed by the detection of an associated broad-lined Type Ic supernova \citep{Sun2024, vanDalen2024, Srivastava2024}, confirming its collapsar origin. Radio analyses constrained the outflow to have at least a moderate $\Gamma$ \citep{Bright2024}, similar to AT 2023sva.

EP and SVOM are opening the door to detecting the prompt emission of events that previously would have been orphan afterglows, enabling a full characterization of their prompt and afterglow emission across the electromagnetic spectrum. Therefore, the coming years hold the tantalizing prospect of breaking historical degeneracies between different GRB models and getting closer to deciphering the landscape of relativistic jets originating from massive stellar deaths. 

\section*{Data Availability}
All of AT 2023sva's optical photometry, radio flux densities, and X-ray upper limits are available in this article. We will make the spectrum available in the supplementary material available online. 

\section*{Acknowledgments}
G.P.S. thanks Isiah Holt for useful discussions on nested sampling techniques, Tony Piro for reading the paper on request, and Simi Bhullar for her moral support through out the paper-writing process. A.Y.Q.H. was supported in part by NASA Grant 80NSSC23K1155. M.W.C acknowledges support from the National Science Foundation with grant numbers PHY-2308862 and PHY-2117997. M.B.S. acknowledges the Finnish Cultural Foundation grant number 00231098 and Finnish Centre for Astronomy with ESO (FINCA) grant. G.C.A. thanks the Indian National Science Academy for support under the INSA Senior Scientist Programme. M.M.K acknowledges generous support from the David and Lucille Packard Foundation. B.O. is supported by the McWilliams Postdoctoral Fellowship at Carnegie Mellon University.

SED Machine is based upon work supported by the National Science Foundation under Grant No. 1106171. Based on observations obtained with the Samuel Oschin Telescope 48-inch and the 60-inch Telescope at the Palomar Observatory as part of the Zwicky Transient Facility project. ZTF is supported by the National Science Foundation under Grant No. AST-2034437 and a collaboration including Caltech, IPAC, the Weizmann Institute of Science, the Oskar Klein Center at Stockholm University, the University of Maryland, Deutsches Elektronen-Synchrotron and Humboldt University, the TANGO Consortium of Taiwan, the University of Wisconsin at Milwaukee, Trinity College Dublin, Lawrence Livermore National Laboratories, IN2P3, University of Warwick, Ruhr University Bochum and Northwestern University. Operations are conducted by COO, IPAC, and UW. The ZTF forced-photometry service was funded under the Heising-Simons Foundation grant \#12540303 (PI: Graham). The Gordon and Betty Moore Foundation, through both the Data-Driven Investigator Program and a dedicated grant,
provided critical funding for SkyPortal. 

The GROWTH India Telescope (GIT, Kumar et al. 2022) is a 70-cm telescope with a 0.7-degree field of view, set up by the Indian Institute of Astrophysics (IIA) and the Indian Institute of Technology Bombay (IITB) with funding from DST-SERB and IUSSTF. It is located at the Indian Astronomical Observatory (Hanle), operated by IIA. We acknowledge funding by the IITB alumni batch of 1994, which partially supports the operations of the telescope. Telescope technical details are available at https://sites.google.com/view/growthindia/. CZTI is built by a TIFR-led consortium of institutes across India, including VSSC, URSC, IUCAA, SAC, and PRL. The Indian Space Research Organisation funded, managed, and facilitated the project. This work is partially based on observations made with the Gran Telescopio Canarias (GTC), installed at the Spanish Observatorio del Roque de los Muchachos of the Instituto de Astrofísica de Canarias, on the island of La Palma. This work was also based on observations made with the Nordic Optical Telescope, owned in collaboration by the University of Turku and Aarhus University, and operated jointly by Aarhus University, the University of Turku and the University of Oslo, representing Denmark, Finland and Norway, the University of Iceland and Stockholm University at the Observatorio del Roque de los Muchachos, La Palma, Spain, of the Instituto de Astrofisica de Canarias. The National Radio Astronomy Observatory is a facility of the National Science Foundation operated under cooperative agreement by Associated Universities, Inc. We thank the staff of the GMRT that made these observations possible. GMRT is run by the National Centre for Radio Astrophysics of the Tata Institute of Fundamental Research. We thank the staff of the Mullard Radio Astronomy Observatory for their invaluable assistance in the operation of the Arcminute Microkelvin Imager.

%%%%%%%%%%%%%%%%%%%% REFERENCES %%%%%%%%%%%%%%%%%%

% The best way to enter references is to use BibTeX:

\bibliographystyle{mnras}
\bibliography{main} % if your bibtex file is called example.bib

% Alternatively you could enter them by hand, like this:
% This method is tedious and prone to error if you have lots of references
%\begin{thebibliography}{99}
%\bibitem[\protect\citeauthoryear{Author}{2012}]{Author2012}
%Author A.~N., 2013, Journal of Improbable Astronomy, 1, 1
%\bibitem[\protect\citeauthoryear{Others}{2013}]{Others2013}
%Others S., 2012, Journal of Interesting Stuff, 17, 198
%\end{thebibliography}

%%%%%%%%%%%%%%%%%%%%%%%%%%%%%%%%%%%%%%%%%%%%%%%%%%

%%%%%%%%%%%%%%%%% APPENDICES %%%%%%%%%%%%%%%%%%%%%

\appendix
\section{Corner Plots}
Here we show the corner plots for the modeling described in \S \ref{modeling}.
\begin{figure*}
    \centering
\includegraphics[width=0.9\linewidth]{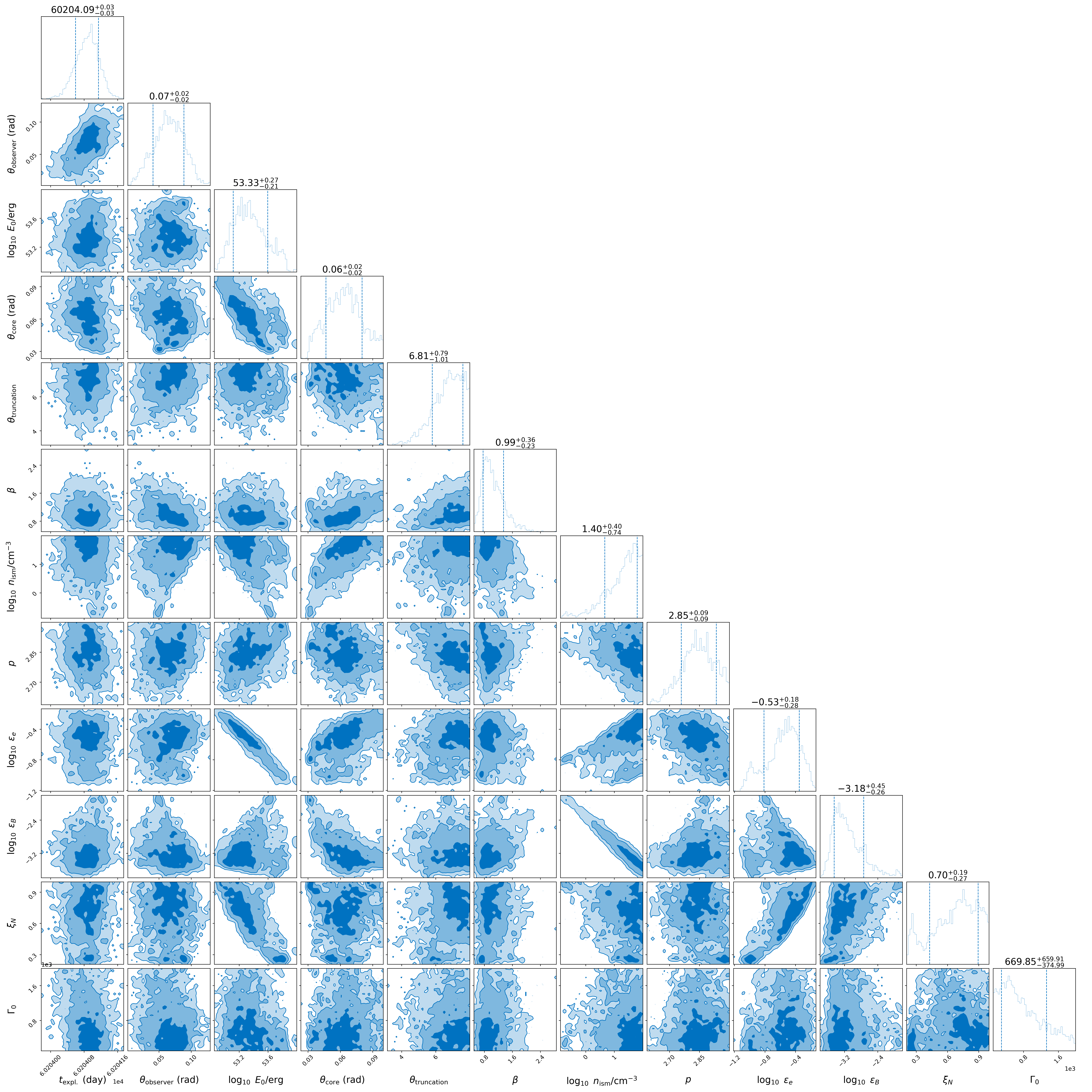}
    \caption{Corner plot for the power-law structured jet model, generated through \texttt{redback}.}
    \label{cornerplot}
\end{figure*}

\begin{figure*}
    \centering
    \includegraphics[width=0.9\linewidth]{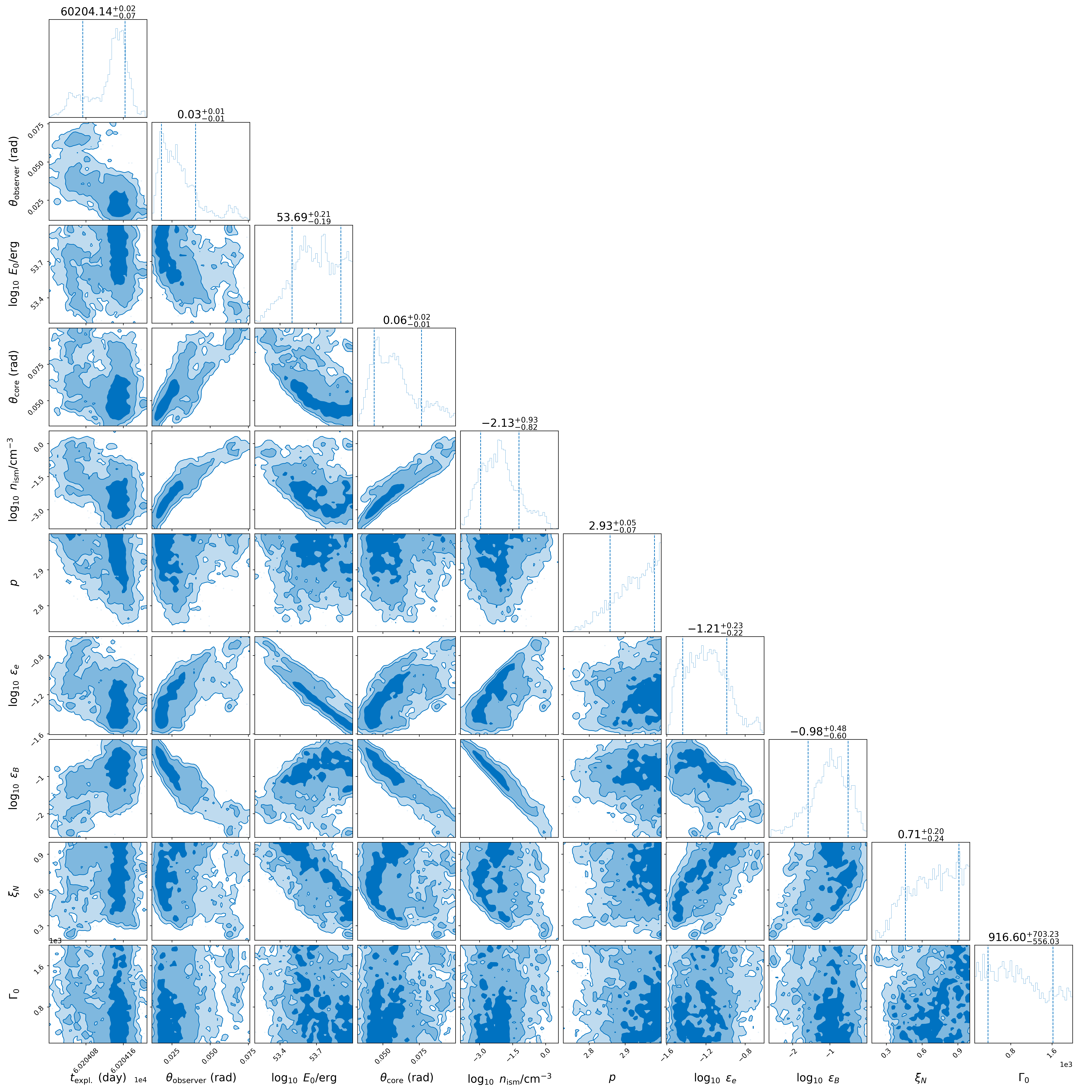}
    \caption{Corner plot for the tophat model, generated through \texttt{redback}.}
    \label{cornerplot}
\end{figure*}

\begin{figure*}
    \centering
    \includegraphics[width=0.9\linewidth]{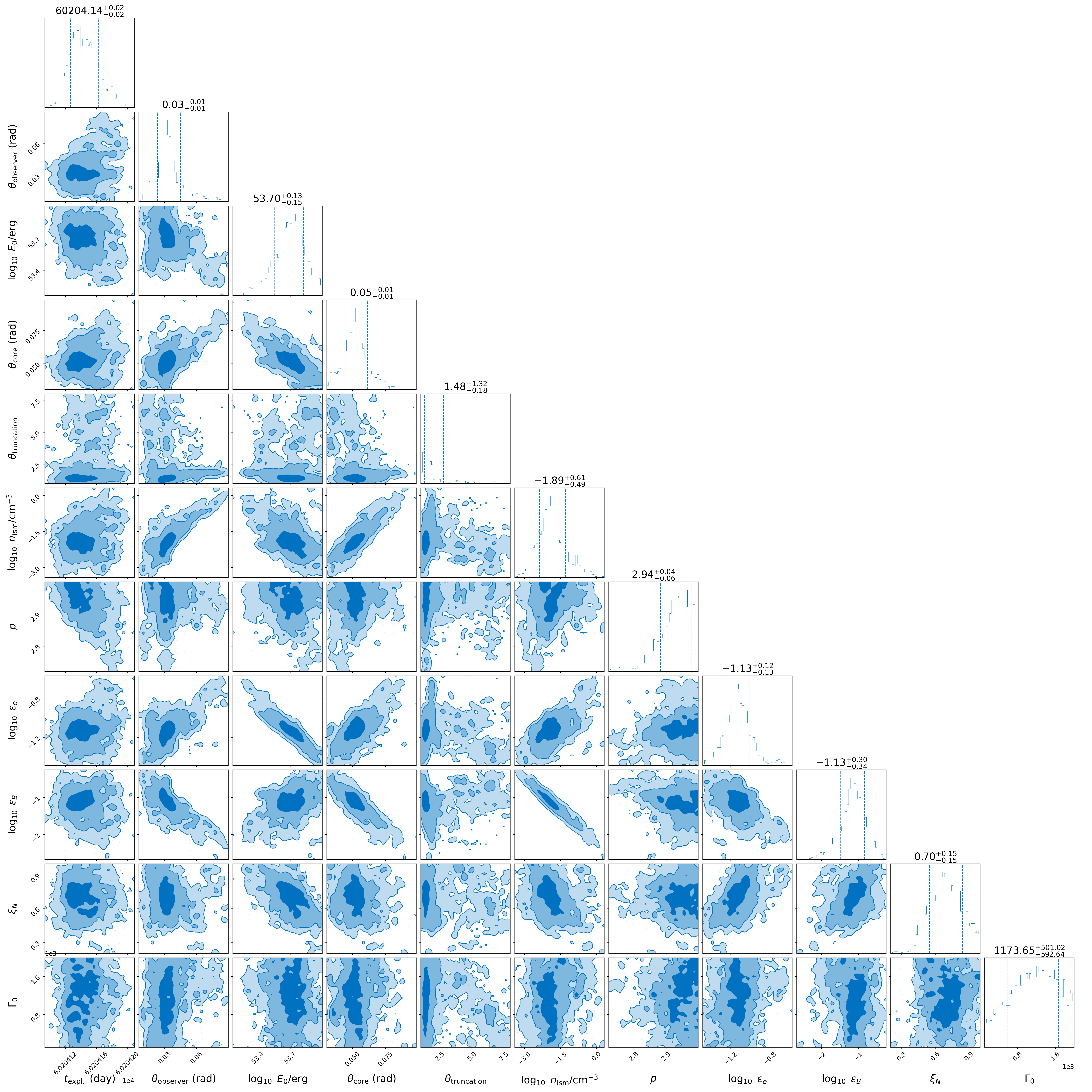}
    \caption{Corner plot for the gaussian structured jet model, generated through \texttt{redback}.}
    \label{cornerplot}
\end{figure*}

%%%%%%%%%%%%%%%%%%%%%%%%%%%%%%%%%%%%%%%%%%%%%%%%%%

% Don't change these lines
\bsp	% typesetting comment
\label{lastpage}
\end{document}